%% file: main.tex
\documentclass[10pt,journal,compsoc]{IEEEtran}

\usepackage{tikz}
\usepackage{amsmath}
\usepackage{url}
\usepackage{balance}
\usepackage{listings}
\usepackage{calc}
\usepackage{bbding}
\usepackage{pifont}
\usepackage{wasysym}
\usepackage{amssymb}
\usepackage{xcolor}
\usepackage{graphicx}
\usepackage{caption,subcaption}
\usepackage{enumitem}
\usepackage{supertabular}
\usepackage{colortbl}
\usepackage{microtype}
\usepackage{fancyvrb}
\usepackage{multicol}
\usepackage{rotating}
\usepackage{enumitem}
\usepackage{xspace}
\usepackage{booktabs}

\usepackage[ruled,vlined,linesnumbered]{algorithm2e}
\usepackage{tikz}
\usepackage{mdframed}
\usepackage{multirow}
\usepackage{bigstrut}
\usepackage{mathtools}
\usepackage{mathpartir}
\usepackage{comment}
\usepackage{tcolorbox}
\usepackage{pifont}

\newtheorem{myDef}{Definition}

\definecolor{lightgray}{rgb}{.9,.9,.9}
\definecolor{darkgray}{rgb}{.4,.4,.4}
\definecolor{purple}{rgb}{0.65, 0.12, 0.82}
\definecolor{pinegreen}{rgb}{0.0, 0.47, 0.44}
\definecolor{bleudefrance}{rgb}{0.19, 0.55, 0.91}
\definecolor{verylightgray}{rgb}{.97,.97,.97}

\lstdefinelanguage{Solidity}{
	keywords=[1]{anonymous, assembly, assert, balance, break, call, callcode, case, catch, class, constant, continue, constructor, contract, debugger, default, delegatecall, delete, do, else, emit, event, experimental, export, external, false, finally, for, function, gas, if, implements, import, in, indexed, instanceof, interface, internal, is, length, library, log0, log1, log2, log3, log4, memory, modifier, new, payable, pragma, private, protected, public, pure, push, require, return, returns, revert, selfdestruct, send, solidity, storage, struct, suicide, super, switch, then, this, throw, transfer, true, try, typeof, using, value, view, while, with, addmod, ecrecover, keccak256, mulmod, ripemd160, sha256, sha3}, 
	keywordstyle=[1]\color{blue}\bfseries,
	keywords=[2]{address, bool, byte, bytes, bytes1, bytes2, bytes3, bytes4, bytes5, bytes6, bytes7, bytes8, bytes9, bytes10, bytes11, bytes12, bytes13, bytes14, bytes15, bytes16, bytes17, bytes18, bytes19, bytes20, bytes21, bytes22, bytes23, bytes24, bytes25, bytes26, bytes27, bytes28, bytes29, bytes30, bytes31, bytes32, enum, int, int8, int16, int24, int32, int40, int48, int56, int64, int72, int80, int88, int96, int104, int112, int120, int128, int136, int144, int152, int160, int168, int176, int184, int192, int200, int208, int216, int224, int232, int240, int248, int256, mapping, string, uint, uint8, uint16, uint24, uint32, uint40, uint48, uint56, uint64, uint72, uint80, uint88, uint96, uint104, uint112, uint120, uint128, uint136, uint144, uint152, uint160, uint168, uint176, uint184, uint192, uint200, uint208, uint216, uint224, uint232, uint240, uint248, uint256, var, void, ether, finney, szabo, wei, days, hours, minutes, seconds, weeks, years},	
	keywordstyle=[2]\color{teal}\bfseries,
	keywords=[3]{block, blockhash, coinbase, difficulty, gaslimit, number, timestamp, msg, data, gas, sender, sig, value, now, tx, gasprice, origin},	
	keywordstyle=[3]\color{violet}\bfseries,
	identifierstyle=\color{black},
	sensitive=false,
	comment=[l]{//},
	morecomment=[s]{/*}{*/},
	commentstyle=\color{gray}\ttfamily,
	stringstyle=\color{red}\ttfamily,
	morestring=[b]',
	morestring=[b]"
}

\newcommand{\slither}{\textsc{Slither}\xspace}
\newcommand{\solhint}{\textsc{Solhint}\xspace}

\newcommand{\oyente}{\textsc{Oyente}\xspace}
\newcommand{\sfuzz}{\textsc{sFuzz}\xspace}
\newcommand{\contractfuzzer}{\textsc{ContractFuzzer}\xspace}
\newcommand{\clairvoyance}{\textsc{Clairvoyance}\xspace}

\newcommand{\smartcheck}{\textsc{SmartCheck}\xspace}
\newcommand{\securify}{\textsc{Securify}\xspace}

\newcommand{\ourTool}{\textsc{xFuzz}\xspace}
\def\WithComments{}
\ifdefined \WithComments
\newcommand{\xyx}[1]{#1}

\newcommand{\revf}[1]{#1}

\else
\newcommand{\xyx}[1]{\textcolor{black}{#1}}

\fi

\newcommand{\codeff}[1]{\texttt{\small #1}}

\definecolor{todocolor}{rgb}{0.9,0.1,0.1}
\definecolor{ylcolor}{rgb}{0.7,0.7,0.3}

\newcommand{\todo}[1]{#1}

\newcommand{\functcall}[1]{\textcircled{\small{#1}}}

\begin{document}

\title{xFuzz: Machine Learning Guided Cross-Contract Fuzzing}

\author{Yinxing~Xue, Jiaming~Ye, Wei~Zhang, Jun~Sun, Lei~Ma, Haijun~Wang, and Jianjun~Zhao
\IEEEcompsocitemizethanks{
\IEEEcompsocthanksitem Yinxing~Xue and Wei~Zhang are with the University of Science and Technology of China. E-mail: yxxue@ustc.edu.cn, sa190@mail.ustc.edu.cn.
\IEEEcompsocthanksitem Jiaming~Ye and Jianjun~Zhao are with the Kyushu University. Email: ye.jiaming.852@s.kyushu-u.ac.jp, zhao@ait.kyushu-u.ac.jp.
\IEEEcompsocthanksitem Jun~Sun is with the Singapore Management University. E-mail: junsun@smu.edu.sg.
\IEEEcompsocthanksitem Lei~Ma is with the University of Alberta. E-mail: ma.lei@acm.org.
\IEEEcompsocthanksitem Haijun~Wang is with the Nanyang Technological University. E-mail: hjwang.china@gmail.com.}
\thanks{Manuscript received December 22, 2021; revised April 14, 2022; accepted June 2, 2022. Date of publication July 2, 2022; date of current version June 5, 2022. This work was supported in part by National Nature Science Foundation of China under Grant 61972373, in part by the Basic Research Program of Jiangsu Province under Grant BK20201192 and in part by the National Research Foundation Singapore under its NSoE Programme (Award Number: NSOE-TSS2019-03). The research of Dr Xue is also supported by CAS Pioneer Hundred Talents Program of China. (Yinxing~Xue and Jiaming~Ye are co-first authors. Yinxing~Xue is the corresponding author).}
}

\input{sec/abs.tex}
\maketitle


\IEEEdisplaynontitleabstractindextext
\IEEEpeerreviewmaketitle

\input{sec/1-intro.tex}
\input{sec/2-background}
\input{sec/3-System}
\input{sec/4-MLmethod}
\input{sec/5-FuzzyTesting}
\input{sec/6-eval}
\input{sec/7-related}
\input{sec/8-conclusion}

\input{sec/ref}

\newpage

\end{document}

%% file: sec/abs.tex
\IEEEtitleabstractindextext{
\begin{abstract}

Smart contract transactions are increasingly interleaved by cross-contract calls. While many tools have been developed to identify a common set of vulnerabilities, the cross-contract vulnerability is overlooked by existing tools.
Cross-contract vulnerabilities are exploitable bugs that manifest in the presence of more than two interacting contracts. Existing methods are however limited to analyze a maximum of two contracts at the same time. Detecting cross-contract vulnerabilities is highly non-trivial. With multiple interacting contracts, the search space is much larger than that of a single contract. To address this problem, we present \ourTool, a machine learning guided smart contract fuzzing framework. The machine learning models are trained with novel features (e.g., word vectors and instructions) and are used to filter likely benign program paths. Comparing with existing static tools, machine learning model is proven to be more robust, avoiding directly adopting manually-defined rules in specific tools. We compare \ourTool with three state-of-the-art tools on \todo{7,391} contracts. \ourTool detects \todo{18} \xyx{exploitable} cross-contract vulnerabilities, of which \todo{15} vulnerabilities are exposed for the first time. Furthermore, our approach is shown to be efficient in detecting non-cross-contract vulnerabilities as well---using less than 20\% time as that of other fuzzing tools, \ourTool detects twice as many vulnerabilities.

\end{abstract}


}

%% file: sec/1-intro.tex
This paper is accepted by IEEE Transactions of Dependable and Secure Computing.

\section{Introduction}
\IEEEPARstart{E}{thereum} has been on the forefront of most rankings of block-chain platforms in recent years~\cite{topplatform}. It enables the execution of programs, called smart contracts, written in Turing-complete languages such as Solidity. Smart contracts are increasingly receiving more attention, e.g., with over 1 million transactions per day since 2018~\cite{ethereum-transaction-rise}.

At the same time, smart contracts related security attacks are on the rise as well. According to \cite{vulnerability-survey-1, vulnerability-survey-2, vulnerability-survey-3}, vulnerabilities in smart contracts have already led to devastating financial losses over the past few years. In 2016, the notorious DAO attack resulted in the loss of 150 million dollars~\cite{DAO-attack}. Additionally, as figured out by Zou \emph{et al.}~\cite{smart-contract-challenges}, over 75\% of developers agree that the smart contract software has a much high security requirement than traditional software. Considering the close connection between smart contract and financial activities, the security of smart contract security largely effects the stability of the society.

 Many methods and tools have since been developed to analyze smart contracts. Existing tools can roughly be categorized into two groups: \emph{static analyzers} and \emph{dynamic analyzers}. Static analyzers (e.g., \cite{oyente, securify, slither, solhint, zeus, smartcheck}) often leverage static program analysis techniques (e.g., symbolic execution and abstract interpretation) to identify suspicious program traces. Due to the well-known limitations of static analysis, there are often many false alarms. On the other side, dynamic analyzers (including fuzzing engines such as~\cite{sfuzz, contractfuzzer, echidna, ethploit, dapp-test}) avoid false alarms by dynamically executing the traces. Their limitation is that there can often be a huge number of program traces to execute and thus smart strategies must be developed to selectively test the program traces in order to identify as many vulnerabilities as possible. Besides, static and dynamic tools also have a common drawback --- \emph{the detection rules are usually built-in and predefined by developers}, sometimes the rules among different tools could be contradictory (e.g., reentrancy detection rules in \slither and \oyente~\cite{clarivoyance}).  



While existing efforts have identified an impressive list of vulnerabilities, one important category of vulnerabilities, i.e., cross-contract vulnerabilities, has been largely overlooked so far. Cross-contract vulnerabilities are exploitable bugs that manifest only in the presence of more than two interacting contracts. For instance, the reentrancy vulnerability shown in Figure \ref{fig:motivation example} occurs only if three contracts interact in a particular order. In our preliminary experiment, the two well-known fuzzing engines for smart contracts, i.e., \contractfuzzer~\cite{contractfuzzer} (version 1.0) and \sfuzz~\cite{sfuzz} (version 1.0), both missed this vulnerability due to they are limited to analyze two contracts at the same time.


Given a large number of cross-contract transactions in practice~\cite{wallet-statistic}, there is an urgent need for developing systematic approaches to identify cross-contract vulnerabilities. Detecting cross-contract vulnerabilities however is non-trivial. With multiple contracts involved, the search space is much larger than that of a single contract, i.e., we must consider all sequences and interleaving of function calls from multiple contracts. 

As fuzzing techniques practically run programs and barely produce false positive reports~\cite{contractfuzzer, harvey}, adopting fuzzing in cross-contract vulnerability detection is preferred. However, due to the efficiency concerns, we need other techniques to guide fuzzers to practically detect cross-contract vulnerabilities. Previous works (e.g., \cite{leopard}, \cite{learntofuzz}) have evidenced the advantages of applying machine learning method for improving efficiency of vulnerability fuzzing in C/C++ programs. Compared with static rule-based methods, the ML model based method requires no prior domain knowledge about known vulnerabilities, and can effectively reduce the large search space for covering more vulnerable functions. In smart contract, existing works (e.g., ILF~\cite{ilf}) focus on exploring the state space in the \emph{intra-contract} scope. They are unable to address the cross-contract vulnerabilities. With a large search space of combinations of numerous function calls, it is desired to guide the fuzzing process via the aid of the machine learning models.



In this work, we propose \ourTool, a machine learning (ML) guided fuzzing engine designed for detecting cross-contract vulnerabilities. Ideally, according to the Pareto principle in testing~\cite{pareto-testing} (i.e., roughly 80\% of errors come from 20\% of the code), \emph{we want to rapidly identify the error-prone code before applying the fuzzing technique}. As reported by previous works~\cite{issta2020smartcontract, issta2021smartcontract}, the existing analysis tools suffer from high false positive rates (e.g., \slither~\cite{slither} and \smartcheck~\cite{smartcheck} have more than 70\% of false positive rates). Therefore, adopting only one static tool in our approach may produce biased results. To alleviate this, we use three tools to vote the reported vulnerabilities in contracts, and we further train a ML model to learn common patterns from the voting results. It is known that ML models can automatically learn patterns from inputs with less bias~\cite{zhuang2020smart}. Based on this, the overall bias due to using a certain tool to identify potentially vulnerable functions in contracts can be reduced.

Specifically, \ourTool provides multiple ways of reducing the enormous search space. First, \ourTool is designed to leverage an ML model for identifying the most probably vulnerable functions. That is, an ML model is trained to filter most of the benign functions whilst preserving most of the vulnerable functions. During the training phase, the ML models are trained based on a training dataset that contains program codes that are labeled using three famous static analysis tools (i.e., the labels are their majority voting result). Furthermore, the program code is vectorized into vectors based on word2vec~\cite{word2vec}. In addition, manually designed features, such as \codeff{can\_send\_eth}, \codeff{has\_call} and \codeff{callee\_external}, are supplied to improve training effectiveness as well. In the guided fuzzing phase, the model is used to predict whether a function is potentially vulnerable or not. In our evaluation of ML models, the models allow us to filter \todo{80.1\%} non-vulnerable contracts.
Second, to further reduce the effort required to expose cross-contract vulnerabilities, the filtered contracts and functions are further prioritized based on a suspiciousness score, which is defined based on an efficient measurement of the likelihood of covering the program paths.

To validate the usefulness of \ourTool, we performed comprehensive experiments, comparing with a static cross-contract detector \clairvoyance~\cite{clarivoyance} and two state-of-the-art dynamic analyzers, i.e., \contractfuzzer~\cite{contractfuzzer} and~\sfuzz, on widely-used open-dataset (\cite{ren2021empirical}, \cite{smartbugs}) and additional \todo{7,391} contracts. The results confirm the effectiveness of \ourTool in detecting cross-contract vulnerabilities, i.e., \todo{18} cross-contract vulnerabilities have been identified. \todo{15} of them are missed by all the tested state-of-the-art tools. We also show that our search space reduction and prioritization techniques achieve high precision and recall. Furthermore, our techniques can be applied to improve the efficiency of detecting intra-contract vulnerabilities, e.g., \ourTool detects twice as many vulnerabilities as that of \sfuzz and uses less than 20\% of time. 

The contributions of this work are summarized as follows.
\begin{itemize}
	\item To the best of our knowledge, we make the first attempts to formulate and detect three common \emph{cross-contract} vulnerabilities, i.e., {reentrancy}, {delegatecall} and {tx-origin}. 
	
	\item We propose a novel ML based approach to significantly reduce the search space for exploitable paths, achieving well-trained ML models with a recall of \todo{95\%} on a testing dataset of \todo{100K} contracts. \revf{We also find that the trained model can cover a majority of reports of other tools.}
	\item We perform a large-scale evaluation and performed comparative studies with state-of-the-art tools. Leveraging the ML models, \ourTool outperforms the state-of-the-art tools by at least \todo{42.8\%} in terms of recall meanwhile keeping a satisfactory precision of \todo{96.1\%}. 
	\item \ourTool also finds \todo{18} cross-contract vulnerabilities. All of them are verified by security experts from our industry partner. We have published the exploiting code to these vulnerabilities on our anonymous website~\cite{xFuzz} for public access.
\end{itemize}


%% file: sec/2-background.tex
\section{Motivation}
\label{sec:background}

In this section, we first introduce three common types of cross-contract vulnerabilities. Then, we discuss the challenges in detecting these vulnerabilities by state-of-the-art fuzzing engines to motivate our work.

\subsection{Problem Formulation and Definition}

In general, smart contracts are compiled into opcodes \cite{solopcodes} so that they can run on EVM. We say that a smart contract is \emph{vulnerable} if there exists a program trace that allows an attacker to gain certain benefit (typically financial) illegitimately. Formally, a vulnerability occurs when there exist dependencies from certain critical instructions (e.g., \codeff{TXORIGIN} and \codeff{DELEGATECALL}) to a set of specific instructions (e.g., \codeff{ADD}, \codeff{SUB} and \codeff{SSTORE}). Therefore, to formulate the problem, we adopt definitions of vulnerabilities from~\cite{securify,sguard}, based on which we define (control and data) dependency and then define the cross-contract vulnerabilities.

\begin{myDef} [\textbf{Control Dependency}] \label{def:control dependency}
An opcode $op_j$ is said to be control-dependent on $op_i$ if there exists an execution from $op_i$ to $op_j$ such that $op_j$ post-dominates all $op_k$ in the path from $op_i$ to $op_j$ (excluding $op_i$) but does not post-dominates $op_i$. An opcode $op_j$ is said to post-dominate an opcode $op_i$ if all traces starting from $op_i$ must go through $op_j$.
\end{myDef} 

\begin{myDef} [\textbf{Data Dependency}] \label{def:data dependency}
An opcode $op_j$ is said to be data-dependent on $op_i$ if there exists a trace that executes $op_i$ and subsequently $op_j$ such that $W(op_i) \cap R(op_j) \neq \emptyset$, where $R(op_j)$ is a set of locations read by $op_j$ and $W(op_i)$ is a set of locations written by $op_i$.
\end{myDef} 

An opcode $op_j$ is \emph{dependent} on $op_i$ if $op_j$ is \emph{control or data dependent} to $op_i$ or $op_j$ is dependent to $op_k$ meanwhile $op_k$ is dependent to $op_i$.

In this work, we define three typical categories of cross-contract vulnerabilities that we focus on, i.e., reentrancy, delegatecall and tx-origin. Although our method can be generalized to support more types of vulnerabilities, in this paper, we focus on the above three vulnerabilities since they are among the most dangerous ones with urgent testing demands. Specifically, the reentrancy and delegatecall vulnerabilities are highlighted as top risky vulnerabilities in previous works~\cite{securify, slither}. The tx-origin vulnerability is broadly warned in previous research~\cite{dasp, slither}.

\begin{figure}
    \centering
    \begin{minipage}[t]{1.0\linewidth}
    \begin{lstlisting}
function withdrawBalance() public {
  uint amountToWithdraw = userBalances[msg.sender];
  msg.sender.call.value(amountToWithdraw)(""); 
  userBalances[msg.sender] = 0;
}
    \end{lstlisting}
    \caption{An example of reentrancy vulnerability.}
    \label{fig:sec2reen-example}
    \end{minipage}
    
    \begin{minipage}[t]{1.0\linewidth}
    \begin{lstlisting}
contract Delegate {
  address public owner;
  function pwn() {
    owner = msg.sender;
  }  }
contract Delegation {
  address public owner;
  Delegate delegate;
  function() {
    if(delegate.delegatecall(msg.data)) {
      this;
  }  }  }
    \end{lstlisting}
    \caption{An example of delegatecall vulnerability.}
    \label{fig:sec2dele-example}	
    \end{minipage}
    
    \begin{minipage}[t]{1.0\linewidth}
    \begin{lstlisting}
function withdrawAll(address _recipient) public {
  require(tx.origin == owner);
  _recipient.transfer(this.balance);
}
    \end{lstlisting}
    \caption{An example of tx-origin vulnerability.}
    \label{fig:sec2tx-example}	
    \end{minipage}
\end{figure}

We define $C$ as a set of critical opcodes, which contains \codeff{CALL}, \codeff{CALLCODE}, \codeff{DELEGATECALL}, i.e., the set of all opcode associated with external calls. These opcodes associated with external calls could be the causes of vulnerabilities (since then the code is under the control of external attackers).

\begin{figure*}[t]
	\centering
	\includegraphics[width=0.8\textwidth]{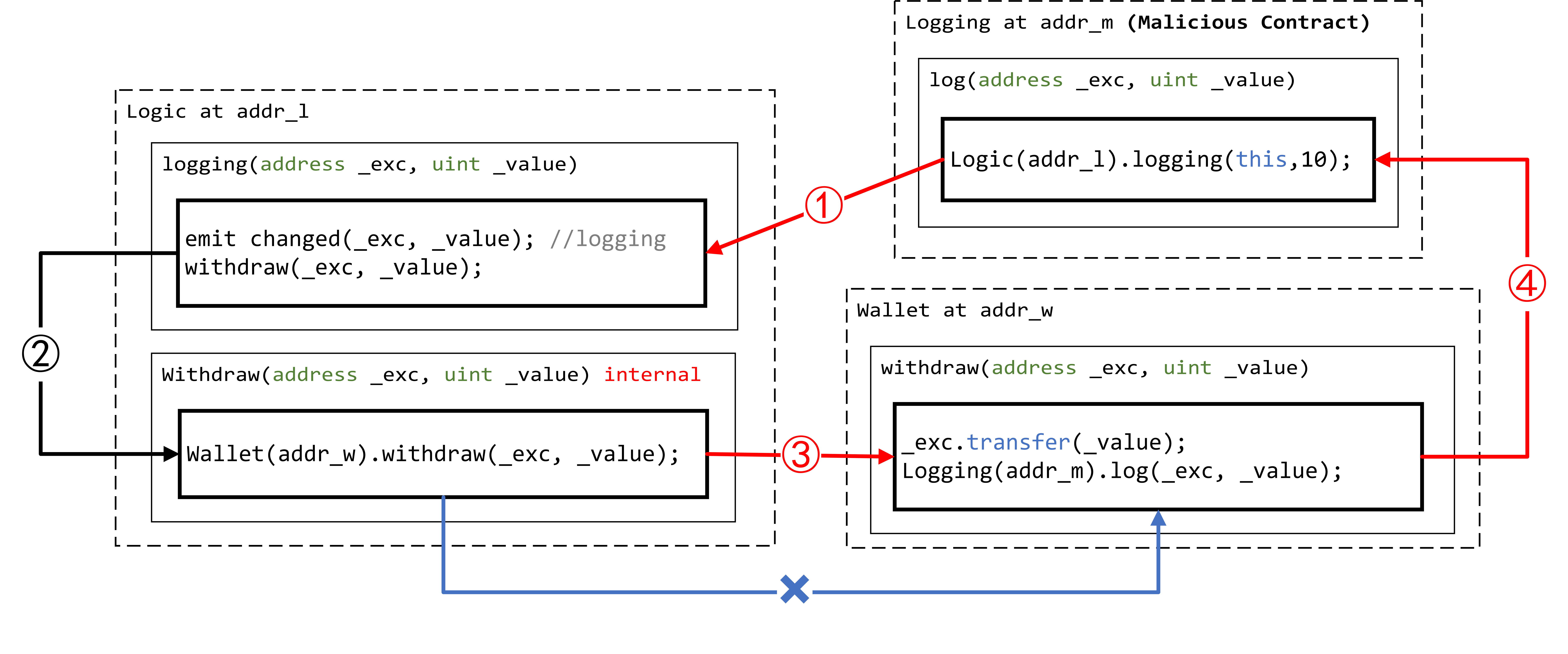}
	\caption{An example of cross-contract reentrancy vulnerability which is missed by the state-of-art fuzzer, namely \sfuzz.}  
	\label{fig:motivation example}
	\begin{minipage}{\textwidth} 
	{\small $^*$Note: The solid boxes represent functions and the dashed containers denote contracts. Specifically, function call is denoted by solid line. The cross-contract calls are highlighted by red arrows. The blue arrow represents cross-contract call missed by sFuzz and ContractFuzzer.}
	\end{minipage}
\end{figure*}
 
\begin{myDef} [\textbf{Reentrancy Vulnerability}] \label{def:reentrancy}
A trace suffers from reentrancy vulnerability if it executes an opcode $op_c \in C$ and subsequently executes an opcode $op_s$ in the same function such that $op_s$ is $\codeff{SSTORE}$, and $op_c$ depends on $op_s$.
\end{myDef}

A smart contract suffers from reentrancy vulnerability if and only if at least one of its traces suffers from reentrancy vulnerability. This vulnerability results from the incorrect use of external calls, which are exploited to construct a call-chain. When an attacker \emph{A} calls a user \emph{U} to withdraw money, the fallback function in contract \emph{A} is invoked. Then, the malicious fallback function calls back to \emph{U} to recursively steal money. In Figure \ref{fig:sec2reen-example}, the attacker can construct an end-to-end call-chain by calling \codeff{withdrawBalance} in the fallback function of the attacker's contract then steals money.

\begin{myDef} [\textbf{Dangerous Delegatecall Vulnerability}] \label{def:delegatecall}
A trace suffers from dangerous delegatecall vulnerability if it executes an opcode $op_c \in C$ that depends on an opcode \codeff{DELEGATECALL}.
\end{myDef}

A smart contract suffers from delegatecall vulnerability if and only if at least one of its traces suffers from delegatecall vulnerability. This vulnerability is due to the abuse of dangerous opcode \codeff{DELEGATECALL}. When a malicious attacker \emph{B} calls contract \emph{A} by using \codeff{delegatecall}, contract \emph{A}'s function is executed in the context of attacker,
and thus causes damages. In Figure \ref{fig:sec2dele-example}, malicious attacker \emph{B} sends ethers to contract \codeff{Delegation} to invoke the fallback function at line 10. The fallback function calls contract \codeff{Delegate} and executes the malicious call data \codeff{msg.data}. Since the call data is executed in the context of \codeff{Delegate}, the attacker can change the owner to an arbitrary user by executing \codeff{pwn} at line 3.

\begin{myDef} [\textbf{Tx-origin Misuse Vulnerability}] \label{def:txorigin}
A trace suffers from tx-origin misuse vulnerability if it executes an opcode $op_c \in C$ that depends on an opcode \codeff{ORIGIN}.
\end{myDef}

A smart contract suffers from tx-origin vulnerability if and only if at least one of its traces suffers from tx-origin vulnerability. This vulnerability is due to the misuse of \codeff{tx.origin} to verify access. An example of such vulnerability is shown in Figure \ref{fig:sec2tx-example}. When a user \emph{U} calls a malicious contract \emph{A}, who intends to forward call to contract \emph{B}. Contract \emph{B} relies on vulnerable identity check (i.e., \codeff{require(tx.origin == owner)} at line 2 to filter malicious access. Since \codeff{tx.orign} returns the address of \emph{U} (i.e., the address of \codeff{owner}), malicious contract \emph{A} successfully poses as \emph{U}. 

\begin{myDef} [\textbf{Cross-contract Vulnerability}] 
A group of contracts suffer from cross-contract vulnerability if there is a vulnerable trace (that suffers from reentrancy, delegatecall, tx-origin) due to opcode from more than two contracts.
\end{myDef}

A smart contract suffers from cross-contract vulnerability if and only if at least one of its traces suffers from cross-contract vulnerability. For example, a cross-contract reentrancy vulnerability is shown in Figure \ref{fig:motivation example}. An attack requires the participation of three contracts: malicious contract \codeff{Logging} deployed at \codeff{addr\_m}, logic contract \codeff{Logic} deployed at \codeff{addr\_l} and wallet contract \codeff{Wallet} deployed at \codeff{addr\_w}. First, the attack function \codeff{log} calls function \codeff{logging} at \codeff{Logic} contract then sends ethers to the attacker contract by calling function \codeff{withdraw} at contract \codeff{Wallet}. Next, the wallet contract sends ethers to attacker contract and calls function \codeff{log}. An end-to-end call chain $\functcall{1} \rightarrow \functcall{2} \rightarrow \functcall{3} \rightarrow \functcall{4} \rightarrow \functcall{1} ...$ is formed and the attacker can recursively steal money without any limitations. 

\subsection{State-of-the-arts and Their Limitations}


First, we perform an investigation on the capability in detecting vulnerabilities by the state-of-the-art methods, including \cite{slither, oyente, securify, clarivoyance, sfuzz, contractfuzzer}. In general, cross-contract testing and analysis are not supported by most of these tools except \clairvoyance. The reason is existing approaches merely focus on one or two contracts, and thus, the sequences and interleavings of function calls from multiple contracts are often ignored. For example, the vulnerability in Figure \ref{fig:motivation example} is a false negative case of static analyzer \slither, \oyente and \securify. Note that although this vulnerability is found by \clairvoyance, this tool however generates many false alarms, making the confirmation of which rather difficult.
This could be a common problem for many static analyzers.


Although high false positive rate could be well addressed by fuzzing tools by running contracts with generated inputs, existing techniques are limited to maximum two contracts (i.e., input contract and tested contract). In our investigation of two currently representative fuzzing tools \sfuzz and \contractfuzzer, cross-contract calls are largely overlooked, and thus leads to missed vulnerabilities. To sum up, most of the existing methods and tools are still limited to handle non-cross-contract vulnerabilities, which motivates this work to bridge such a gap towards solving the currently urgent demands.

%% file: sec/3-System.tex
\begin{figure*}[t]
\setlength{\belowcaptionskip}{-0.45cm}
	\centering
	\includegraphics[width=\textwidth]{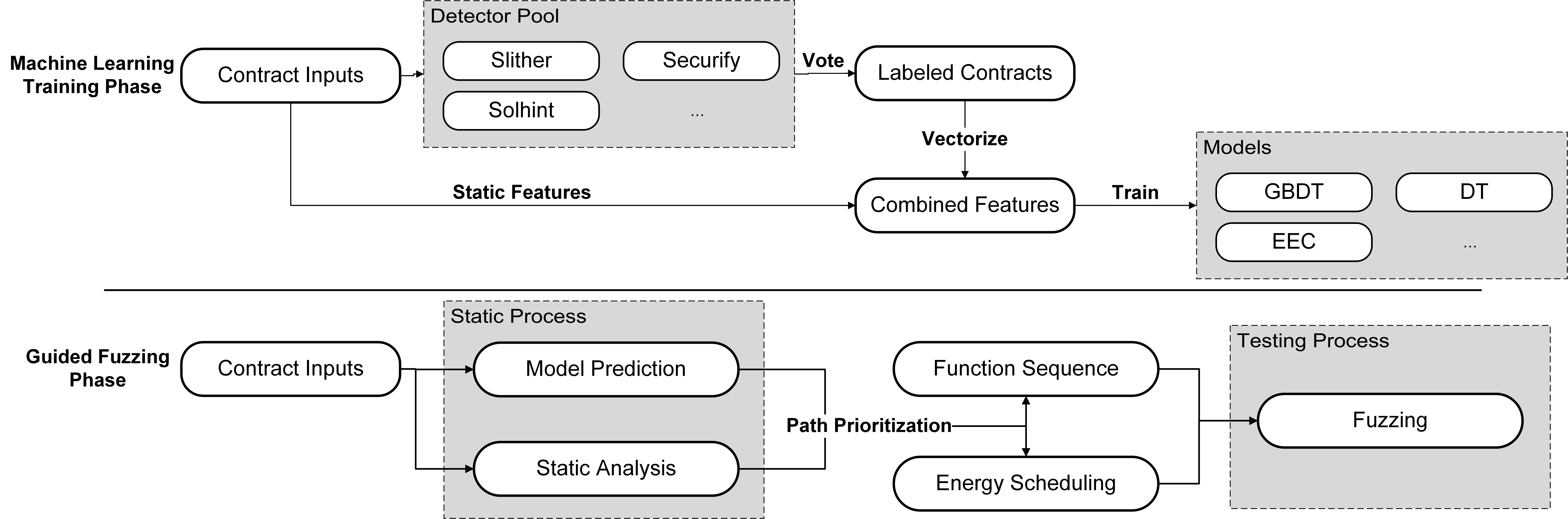}
	\caption{The overview of \ourTool framework.}  \label{fig:system_diagram}
	\medskip 
\end{figure*}

\section{Overview}
\label{sec:overview}

Detecting cross-contract vulnerability often requires examining a large number of sequence transactions and thus can be quite computationally expensive some even infeasible. In this section, we give an overall high-level description of our method, e.g., focusing on fuzzing suspicious transactions based on the guideline of a machine learning (ML) model. Technically, there are three challenges of leveraging ML to guide the effective fuzzing cross-contracts for vulnerability detection:

\begin{itemize}
\item[\textbf{C1}] How to train the machine learning model and achieve \emph{satisfactory} precision and recall.
\item[\textbf{C2}] How to combine trained model with fuzzer to reduce search space towards \emph{efficient} fuzzing.
\item[\textbf{C3}] How to empower the guided fuzzer the support of \emph{effective} cross-contract vulnerability detection.
\end{itemize}

In the rest of this section, we provide an overview of \ourTool which aims at addressing the above challenges, as shown in Figure~\ref{fig:system_diagram}. Generally, the framework can be separated into two phases: \emph{machine learning model training phase} and \emph{guided fuzzing phase}. 


\subsection{Machine Learning Model Training Phase}

In previous works \cite{driller, flayer}, fuzzers 
are limited to prior knowledge of vulnerabilities and they are not well generalized against vulnerable variants. In this work, we propose to leverage ML predictions to guide fuzzers. The benefit of using ML instead of a particular static tool is that ML model can reduce bias introduced by manually defined detection rules. 


In this phase, we collect training data, engineer features, and evaluate models. First, we employ the state-of-the-arts \slither, \securify and \solhint to detect vulnerabilities on the dataset. Next, we collect their reports to label contracts. The contract gains at least two votes are labeled as vulnerability. After that, we engineer features. The input contracts are compiled into bytecode then vectorized into vectors by Word2Vec \cite{word2vec}. To address \textbf{C1}, they are enriched by combining with static features (e.g., \codeff{can\_send\_eth}, \codeff{has\_call} and \codeff{callee\_external}, etc.). These static features are extracted from ASTs and CFGs. Eventually, the features are used as inputs to train the ML models. In particular, the precision and recall of models are evaluated to choose three candidate models (e.g., XGBoost~\cite{xgboost}, EasyEnsembleClassifier~\cite{easyensemble} and Decision Tree), among which we select the best one. 


\subsection{Guided Testing Phase}


In guided testing phase, contracts are input to the pretrained models to obtain predictions. After that, the vulnerable contracts are analyzed and pinpointed. To address challenge \textbf{C2}, the functions that are predicted as suspiciously vulnerable ones. Then we use call-graph analysis and control-flow-graph analysis to construct cross-contract call path. After we collect all available paths, we use the path prioritization algorithm to prioritize them. The prioritization becomes the guidance of the fuzzer. This guidance of model predictions significantly reduces search space because the benign functions wait until the vulnerable ones finish. The fuzzer can focus on vulnerable functions and report more vulnerabilities.
 

To address \textbf{C3}, we extract static information (e.g., function parameters, conditional paths) of contracts to enrich model predictions. The predictions and the static information are combined to compute path priority scores. Based on this, the most exploitable paths are prioritized, where vulnerabilities are more likely found. Here, the search space of exploitable paths is further reduced and the cross-contract fuzzing is therefore feasible by invoking vulnerability through available paths. 


%% file: sec/4-MLmethod.tex
\section{Machine Learning Guidance Preparation}
\label{sec:machine learning method}
In this section, we elaborate on the training of our ML model for fuzzing guidance. We discuss the data collection in Section \ref{sec:data_collection} and introduce feature engineering in Section \ref{sec:feature_engineering}, followed by candidate model evaluation in Section \ref{sec:model_selection}.

\subsection{Data Collection}
\label{sec:data_collection}

\textsc{SmartBugs} \cite{smartbugs} and \textsc{SWCRegistry} \cite{swcregis} are two representatives of existing smart contract vulnerability benchmarks. However, their labeled data is scarce and the amount currently available is insufficient to train a good model. Therefore, we choose to download and collect contracts from Etherscan (https://etherscan.io/), a prominent Ethereum service platform. Overall, to be representative, we collect a large set of \todo{100,139} contracts in total for further processing.

\begin{table}[t]
    \centering
    \caption{Vulnerability detection capability of voting static tools.}
    \label{tbl:detection-capability}
\begin{tabular}{cccc}
\toprule
  & Slither & Solhint & Securify \\
\midrule
Reentrancy  & \ding{108} & \ding{108} & \ding{108} \\
Tx-origin & \ding{108} & \ding{108} & \\
Delegatecall & \ding{108} &  &  \\
\bottomrule
\end{tabular}
\end{table}

The collected dataset is then labeled based on the voting results of three most well-rated static analyzers (i.e., \solhint \cite{solhint} v2.3.1, \slither \cite{slither} v0.6.9 and \securify \cite{securify} v1.0 ). The three tools are chosen based on the fact that they are \ding{182} state-of-the-art static analyzers and \ding{183} well maintained and frequently updated. The detection capability vary among these tools (as shown in Table~\ref{tbl:detection-capability}). We then vote to label the dataset aiming at eliminating the bias of each tool. 
Note that the two vulnerabilities (i.e., delegatecall and tx-origin) are hardly supported by existing tools. Therefore, we only vote vulnerable functions on vulnerabilities supported by at least two tools. That is, for reentrancy, the voting results are counted in the way that the function gain at least two votes is deemed as vulnerability; for tx-origin, the function is deemed as vulnerability when it gains at least one vote. As for delegatecall vulnerability, we label all reported functions as vulnerable ones.

As a result, we collect \todo{788} reentrancy, \todo{40} delegatecall and \todo{334} tx-origin vulnerabilities, respectively. All of the above vulnerabilities are manually confirmed by two authors of this paper, both of whom have more than 3 years development experience for smart contracts, to remove false alarms.


\subsection{Feature Engineering}
\label{sec:feature_engineering}

Then, both vulnerable and benign functions are preprocessed by \slither to extract their runtime bytecode. After that, Word2Vec \cite{word2vec} is leveraged to transform the bytecode into a 20-dimensional vector. However, as reported in \cite{code2vec}, vectors alone are still insufficient for training a high-performance model. To address this, we enrich the vectors with 7 additional static features extracted from CFGs. In short, the features are 27 dimensions in total, in which 20 are yielded by Word2Vec and the other 7 are summarized in Table \ref{tbl:model_features}.

\input{table/features}


Among the 7 static features, \codeff{has\_modifier}, \codeff{has\_call}, \codeff{has\_balance}, \codeff{callee\_external} and \codeff{can\_send\_eth} are static features. We collect them by utilizing static analysis techniques. The feature \codeff{has\_modifier} is designed to identify existing program guards. In smart contract programs, the function modifier is often used to guard a function from arbitrary access. That is, a function with modifier is less like a vulnerable one. Therefore, we make the modifier as a counter-feature to avoid false alarms. Feature \codeff{has\_call} and feature \codeff{has\_balance} are designed to identify external calls and balance check operations. These two features are closely connected with transfer operations. We prepare them to better locate the transfer behavior and narrow search space. Feature \codeff{callee\_external} provides important information on whether the function has external callees. This feature is used to capture risky calls. In smart contracts, cross-contract calls are prone to be exploited by attackers. Feature \codeff{can\_send\_eth} extracts static information (e.g., whether the function has transfer operation) to figure out whether the function has ability to send ethers to others. Considering the vulnerable functions often have risky transfer operations, this feature can help filtering out benign functions and reduce false positive reports.

The remaining three features, i.e., \codeff{has\_delegate} and \codeff{has\_tx\_origin} correspond to particular key opcodes used in vulnerabilities. Specifically, feature \codeff{has\_delegate} corresponds to the opcode \codeff{DELEGATECALL} in delegatecall vulnerabilities, feature \codeff{has\_tx\_origin} corresponds to the opcode \codeff{ORIGIN} in tx-origin vulnerabilities. These two features are specifically designed for the two vulnerabilities, as their names suggest. Note that the features can be easily updated to support detection on new vulnerabilities. If the new vulnerability shares similar mechanism with the above three vulnerabilities or is closely related to them, the existing features can be directly adopted; otherwise, one or two new specific features highly correlated with the new type of vulnerability should be added. The 7 static features are combined with word vectors, which together form the input to our ML models for further training.

\subsection{Model Selection}
\label{sec:model_selection}

In this section, we train and evaluate diverse candidate models, based on which we select the best one to guide fuzzers. To achieve this, one challenge we have to address first is the dataset imbalance. In particular, there are \todo{1,162} vulnerabilities and \todo{98,977} benign contracts. This is not rare in ML-based vulnerability detection tasks~\cite{imbalance-survey-1, imbalance-survey-2}. In fact, our dataset endures imbalance in rate of \todo{1:126} for reentrancy, \todo{1:2,502} for delegatecall and \todo{1:298} for tx-origin. Such imbalanced dataset can hardly be used for training.

To address the challenge, we first eliminate the duplicated data. In fact, we found \todo{73,666} word vectors are exactly same to others. These samples are different in source code, but after they are compiled, extracted and transformed into vectors, they share the same values, because most of them are syntactically identical clones \cite{ccfinder} at source code level. After our remedy, data imbalance comes to \todo{1:31} for reentrancy, \todo{1:189} for delegatecall and \todo{1:141} for tx-origin. Still the dataset is highly imbalanced.

\begin{figure}[t]
	\centering
	\includegraphics[scale=0.55]{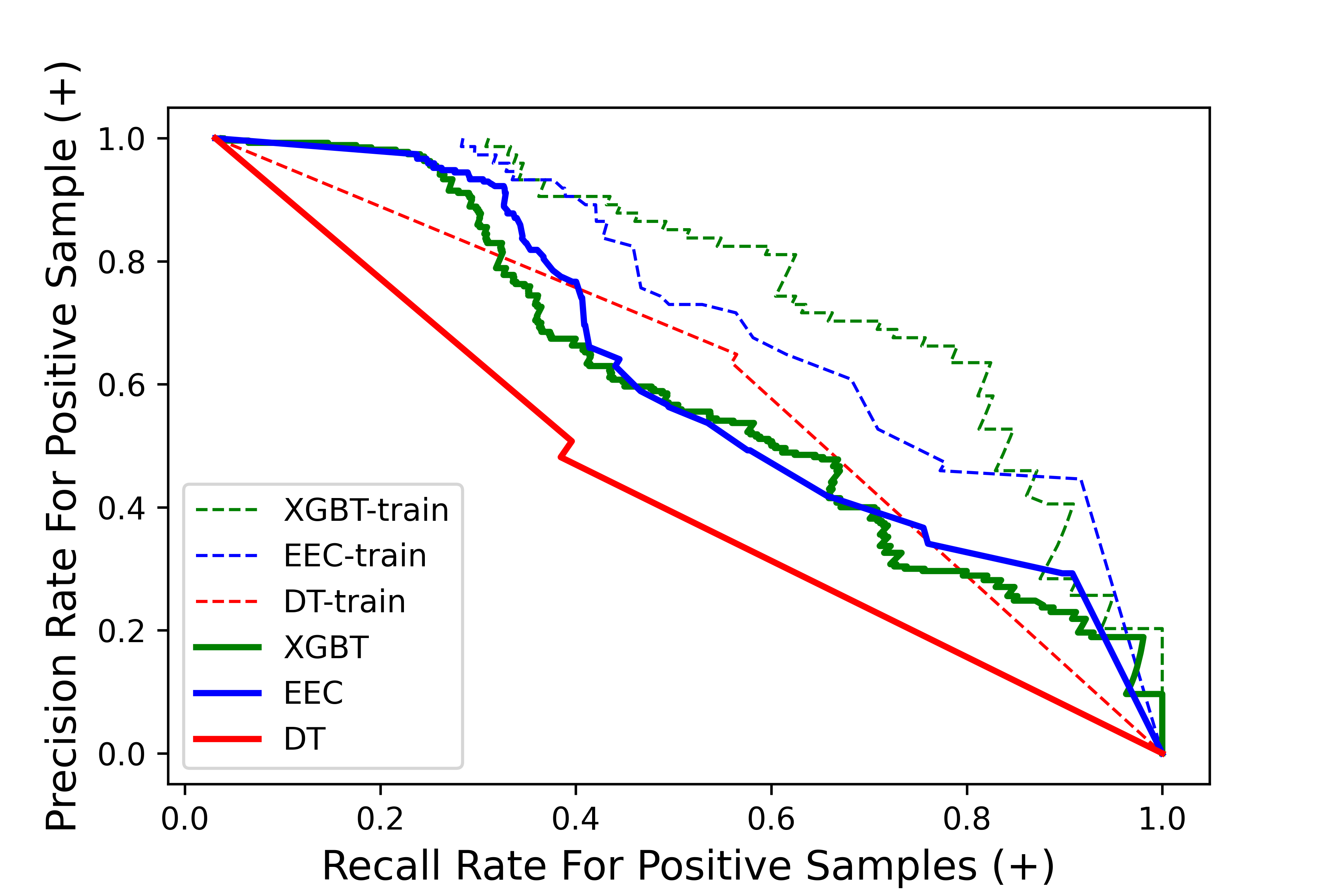}
	\caption{The P-R Curve of models. The dashed lines represent performance on training set, while the solid lines represent performance on validation set.}  
	\label{fig:prcurve}
\end{figure}

As studied in \cite{imbalance-survey}, the imbalance can be alleviated by data sampling strategies. However, we find that sampling strategies like oversampling \cite{smote} can hardly improve the precision and recall of models because the strategy introduces too much polluted data instead of real vulnerabilities. 

\begin{table}[t]
\caption{The performance of evaluated ML models.}
\centering
\label{tbl:mlcompare}
\begin{tabular}{ccc}
\toprule
Model Name & Precision & Recall \\
\midrule
EasyEnsembleClassifier & 26\% & 95\% \\
XGBoost & 66\% & 48\% \\
DecisionTree & 70\% & 43\% \\
SupportVectorMachine & 60\% & 14\% \\
KNeighbors & 50\% & 43\% \\
NaiveBayes & 50\% & 59\% \\
LogiticRegression & 53\% & 38\% \\
\bottomrule
\end{tabular}
\end{table}

We then attempt to evaluate models to select one that fits the imbalanced data well. Note that to counteract the impact of different ML models, we try to cover as many candidate ML methods as possible, among which we select the best one. The models we evaluated including tree-based models XGBT \cite{xgboost}, EEC \cite{easyensemble}, Decision Tree (DT), and other representative ML models like Logistic Regression, Bayes Models, SVMs and LSTM \cite{bilstm}. The performance of the models can be found at Table \ref{tbl:mlcompare}. We find that the tree-based models achieve better precision and recall than others. Other non-tree-based models are biased towards the major class and hence show very poor classification rates on minor classes. Therefore, we select XGBT, EEC and DT as the candidate models.

The precision-recall curves of the three models on positive cases are shown on Figure \ref{fig:prcurve}. In this figure, the dashed lines denote models fitting with validation set and solid lines denote fitting with testing set. Intuitively, model XGBT and model EEC achieve better performance with similar P-R curves. However, EEC performs much better than XGBT in recall. In fact, model XGBT holds a precision rate of \todo{66\%} and a recall rate of \todo{48\%}. Comparatively, model EEC achieves a precision rate of 26\% and a recall rate of 95\%. We remark that our goal is not to train a model that is very accurate, but rather a model that allows us to filter as many benign contracts as possible without missing real vulnerabilities. Therefore, we select the EEC model for further guiding the fuzzing process.

\subsection{Model Robustness Evaluation}

To further evaluate the robustness of our selected model and to assess that to how much extent can our model represent existing analyzers, we conduct evaluation of comparing the vulnerability detection on unknown dataset between our model and other state-of-the-art static analyzers. The evaluation dataset is download from a prominent third-party blockchain security team (https://github.com/tintinweb/smart-contract-sanctuary). We select smart contracts released in version 0.4.24 and 0.4.25 (i.e., the majority versions of existing smart contract applications~\cite{solidityversion}) and remove the contracts which has been used in our previous model training and model selection. After all, we get 78,499 contracts in total for evaluation.



\begin{table}[]
\caption{The coverage rate (CR) score of ML model on other tools.}
\centering
\label{tbl:mlrobust}
\begin{tabular}{cccc}
\toprule
           & CR(Slither) & CR(Securify) & CR(Solhint) \\
\midrule
Reentrancy & 83.6\%   & 81.1\%    & 86.3\%  \\
Tx-origin  & 91.9\%   &    N.A.      & 75.1\%  \\
Delegatecall & 90.6\% & N.A. & N.A. \\
\bottomrule
\end{tabular}
\end{table}

\begin{myDef} [Coverage Rate of ML Model on Another Tool]
\label{def:coverage_rate}
    Given the true positive reports of ML model $R_m$, the true positive reports of another tool $R_t$, a coverage rate of ML model $CR(t)$ on the tool is calculated as:
    \begin{equation}
        CR(t) = (R_m \cap R_t) / R_t
    \end{equation}
\end{myDef} 

The results are listed in Table~\ref{tbl:mlrobust}. Here, we use the coverage rate ($CR$) to evaluate the representativeness of our model regarding the three vulnerabilities. Specifically, the coverage rate measures how much reports of ML model are intersected with static analyzer tools. The coverage rate $CR$ is calculate as listed in Definition~\ref{def:coverage_rate}. The N.A. in the table denotes that the detection of this vulnerability is not support by the analyzer.


Our evaluation results show that the reports of our tool can cover a majority of reports of other tools. Specifically, the trained ML model can well approximate the capability of each static tool used in vulnerability labeling and model training. For example, 81.1\% of true positive reports of \securify on reentrancy are also contained in our ML model’s reports. Besides, 75.1\% of true positive reports of \solhint on Tx-origin and 90.6\% of true positive reports of \slither on Delegatecall are also covered. 


%% file: table/features.tex
\small
\begin{table}[t]
    \centering
    \caption{The seven static features adopted in model training }
    \label{tbl:model_features}
    \begin{tabular}{ccc}
\toprule
 \textbf{Feature Name}  & \textbf{Type} & \textbf{Description}\\ 
\midrule
 has\_modifier  & bool & whether has a modifier \\
 has\_call & bool & whether contains a call operation \\
 has\_delegate & bool & whether contains a delegatecall \\
 has\_tx\_origin & bool & whether contains a tx-origin operation \\
 has\_balance & bool & whether has a balance check operation \\
  can\_send\_eth & bool & whether supports sending ethers \\
 callee\_external  & bool & whether contains external callees \\
\bottomrule
\end{tabular}
\end{table}
\normalsize


%% file: sec/5-FuzzyTesting.tex
\section{Guided Cross-contract Fuzzing}
\label{sec:guided fuzzing}


\begin{algorithm}[t]
	\small
	\SetKwInOut{Input}{input}
	\SetKwInOut{Output}{output}
	\Input{$IS$, all the input smart contract source code}
	\Input{$M$, suspicious function detection ML model}
	\Input{$TRs\leftarrow \emptyset$, the set of potentially vulnerable function execution paths}
	\Output{$V \leftarrow \emptyset$, the set of vulnerable paths}
	$F_s \leftarrow IS.getFunctionList()$ \\ \label{algo:line1}
	// \texttt{\small get the functions in a contract} \\
	\ForEach{function $f \in F_s$}{ \label{algo:line3}
	    \uIf{$ifIsSuspiciousFunction(f, M)$ is True}{
	        // \texttt{\small employ ML models to predict whether the function is suspicious} \\
	        $S_{func} \leftarrow getFuncPriorityScore(f)$ \\
	        $S_{caller} \leftarrow getCallerPriorityScore(f)$ \\
	        $TRs \leftarrow TRs \cup \{f, S_{func}, S_{caller}\}$ \\ \label{algo:line8}
	        // \texttt{get scores for each function} \\ 
	    }
	}
	$PTR \leftarrow PrioritizationAlgorithm(TRs)$ \\ \label{algo:line10}
	// \texttt{\small Prioritized paths} \\
	$V \leftarrow \emptyset$ \\
	// \texttt{\small the output vulnerability list} \\
	\While{not timeout}{ \label{algo:line13}
	    $T \leftarrow PTR.pop()$ \\
	    // \texttt{\small pop up trace with higher priority} \\
	    $FuzzingResult \leftarrow Fuzzing(T)$ \\
	    \eIf{$FuzzingResult$ is Vulnerable} {
	        $V \leftarrow V \cup \{T\}$ \\
	    }{
	        \textbf{continue} \\ \label{algo:line20}
	    }
	}
	\textbf{return} $V$
	\caption{Machine learning guided fuzzing}\label{algo:guided_fuzzing}
\end{algorithm}


\subsection{Guidance Algorithm}

The pretrained models are applied to guide fuzzers in the ways that the predictions are utilized to \ding{182} locate suspicious functions and \ding{183} combine with static information for path prioritization. 


\begin{figure}
    \centering
    \begin{minipage}[tb]{1.0\linewidth}
    \begin{lstlisting}
contract Wallet{
  function withdraw(address addr, uint value){
    addr.transfer(value);
  }
  function changeOwner(address[] addrArray, uint idx) public{
    require(msg.sender == owner);
    owner = addrArray[idx];
    withdraw(owner, this.balance);
  }  }
contract Logic{
  function logTrans(address addr_w, address _exec, uint _value, bytes infor) public{
    Wallet(addr_w).withdraw(_exec, _value);
  }  }
    \end{lstlisting}
    \caption{An example of prioritizing paths.}
    \label{fig:sec5example}	
    \end{minipage}
\end{figure}


Our guidance is based on both model predictions and the priority scores computed from static features. The reason is that even with the machine learning model filtering, the search space is still rather large, which is evidenced by the large number of paths explored by \sfuzz (e.g., the \todo{2,596} suspicious functions have \todo{873} possibly vulnerable paths), and thus we propose to first prioritize the path. 

The overall process of our guided fuzzing can be found at Algorithm~\ref{algo:guided_fuzzing}. In this algorithm, we first retrieve function list of an input source at line~\ref{algo:line1}. Next, from line~\ref{algo:line3} to line~\ref{algo:line8}, we calculate the path priority based on two scores (i.e., function priority scores and caller priority scores) for each path. Both scores are designed for prioritizing suspicious functions. After the calculation, the results are saved together with the function itself. In line~\ref{algo:line10}, we prioritize the suspicious function paths. The prioritization algorithm can be found at Algorithm~\ref{algo2:prioritization}. The trace with higher priority will be first tested by fuzzers. Finally, from line~\ref{algo:line13} to line~\ref{algo:line20}, we pop up a candidate trace from prioritized list and employ fuzzers to conduct focus fuzzing. The fuzzing process will not end until it reaches an timeout limitation. The found vulnerability will be return as final result.

The details of our prioritization algorithm are shown in Algorithm~\ref{algo2:prioritization}. The input of the algorithm is the functions and their corresponding priority scores. The scores are calculated in Algorithm~\ref{algo:guided_fuzzing}. The output of the algorithm is the prioritized vulnerable paths. Specifically, the first step of the algorithm is getting the prioritized function based on the function priority score, as shown in line 2 and line 3. The functions with lower function priority scores will be prioritized. Next, we sort all call paths (no matter cross-contract or non-cross-contract call) which are correlated to the function, as shown from line 4 to line 6. We pop up the call path which has the highest priority and add it to the prioritized path set. The prioritized path set will guide fuzzer to test call path in a certain order.

To summarize, the goal of our guidance algorithm is to prioritize cross-contract paths, which are penetrable but usually overlooked by previous practice~\cite{contractfuzzer, sfuzz}, and to further improve the fuzzing testing efficiency on cross-contract vulnerabilities.


\begin{algorithm}[t]
	\small
	\SetKwInOut{Input}{input}
	\SetKwInOut{Output}{output}
    \Input{$M$, The trained machine learning model}
	\Input{$TRs$, functions and their priority scores}
	\Output{$PTR$, the set of prioritized vulnerable paths}
    \While {$isNotEmpty(TRs)$}{
        $TRs \leftarrow sortByFunctionPriority(TRs)$ \\
        function $f \leftarrow TRs.pop()$ \\
        paths $Ps \leftarrow getAllPaths(f)$ \\
        \While {$isNotEmpty(Ps)$}{
            $Ps \leftarrow sortByCallerPriority(Ps)$ \\
            $P \leftarrow Ps.pop()$ \\
            $PTR \leftarrow PTR \cup P$
        }
    }
    \textbf{return} $PTR$
	\caption{Priorization Algorithm}\label{algo2:prioritization}
\end{algorithm}

\subsection{Priority Score}

Generally, the path priority consists of two parts: \emph{function priority} and \emph{caller priority}. The function priority is for evaluating the complexity of function and the caller priority is designed to measure the cost to traverse a path.

\textbf{Function Priority.} We collect static features of functions to compute function priority. After that, a priority score can be obtained. The lower score denotes higher priority. 

We first mark the suspicious functions by model predictions. A suspicious function is likely to contain vulnerabilities so it is provided with higher priority. We implement this as a factor $f_s$ which equals 0.5 for suspicious function otherwise 1 for benign functions. For example, in Figure \ref{fig:sec5example}, the function \codeff{withdraw} is predicted as suspicious so that the factor $f_s$ equals 0.5.

Next, we compute the caller dimensionality $S_C$. The dimensionality is the number of callers of a function. In cross-contract fuzzing, a function with multiple callers requires more testing time to traverse all paths. For example, in Figure \ref{fig:sec5example}, function \codeff{withdraw} in contract \codeff{Wallet} has an internal caller \codeff{changeOwner} and an external caller \codeff{logTrans}, thus the dimensionality of this function is 2.

The parameter dimensionality $S_P$ is set to measure the complexity of parameters. The functions with complex parameters (i.e., array, bytes and address parameters) are assigned with lower priority, because these parameters often increase the difficulty of penetrating a function. Specifically, one parameter has 1 dimensionality except for the complex parameters, i.e., they have 2 dimensionalities. The parameter dimensionality of a function is the sum of parameters dimensionalities. For example, in Figure \ref{fig:sec5example}, function \codeff{withdraw} and \codeff{changeOwner} both have an address and an integer parameter thus their dimensionality is 3. Function \codeff{logTrans} has two addresses, a byte and an integer parameter, so the dimensionality is 7. 

\begin{myDef} [Function Priority Score]
\label{def:function priority score}
    Given the suspicious factor $f_s$, the caller dimensionality score $S_C$ and the parameter dimensionality score $S_P$, a function priority score $S_{func}$ is calculated as:
    \begin{equation}
        S_{func} = f_s \times (S_C + 1) \times (S_P + 1)
    \end{equation}
\end{myDef} 

In this formula, we add 1 to the caller dimensionality and parameter dimensionality to avoid the overall score to be 0. The priority scores in Figure \ref{fig:sec5example} are: function \codeff{withdraw} = 6, function \codeff{changeOwner} = 4, function \codeff{logTrans} = 8. The results show that function \codeff{changeOwner} has highest priority because function \codeff{withdraw} has two callers to traverse meanwhile function \codeff{logTrans} is more difficult for penetration than \codeff{changeOwner}.

\textbf{Caller Priority.} We traverse every caller of a function and collect their static features, based on which we compute the priority score to decide which caller to test first. Firstly, the number of branch statements (e.g., \codeff{if}, \codeff{for} and \codeff{while}) and assertions (e.g., \codeff{require} and \codeff{assert}) are counted to measure condition complexity $Comp$ to describe the difficulties to bypass the conditions. The path with more conditions is in lower priority. For example, in Figure \ref{fig:sec5example}, function \codeff{withdraw} has two callers. One caller \codeff{changeOwner} has an assertion at line 6, so the complexity is 1. The other caller \codeff{logTrans} contains no conditions thus the complexity is 0.

Next, we count the condition distance.
\sfuzz selects seed according to branch-distance only, which is not ideal for identifying the three particular kinds of cross-contract vulnerabilities that we focus on in this work. Thus, we propose to consider not only branch distance but also this condition distance $CondDis$. 
This distance is intuitively the number of statements from entry to condition. In case of the function has more than one conditions, the distance is the number of statements between entry and first condition. For example, in Figure \ref{fig:sec5example}, the condition distance of \codeff{changeOwner} is 1 and the condition distance of \codeff{logTrans} is 0.

\begin{myDef} [Caller Priority Score]
\label{def:caller priority score}
    Given the condition distance $CondDis$ and the path condition complexity $Comp$, a path priority score $S_{caller}$ is calculated as:
    \begin{equation}
        S_{caller} = (CondDis + 1) \times (Comp + 1)
    \end{equation}
\end{myDef}

Finally, the caller priority score is computed based on condition complexity and condition distance, as shown in Definition \ref{def:caller priority score}. The complexity and distance add 1 so that the overall score is not 0. The caller priority scores in Figure \ref{fig:sec5example} are: \codeff{logTrans} $\rightarrow$ \codeff{withdraw} = 1, \codeff{changeOwner} $\rightarrow$ \codeff{withdraw} = 4. Function \codeff{changeOwner} has identity check at line 6, which increase the difficulty to penetrate. Thus, the other path from \codeff{logTrans} to \codeff{withdraw} is prior.

\subsection{Cross-contract Fuzzing}

\begin{figure}[t]
	\centering
	\includegraphics[scale=0.38]{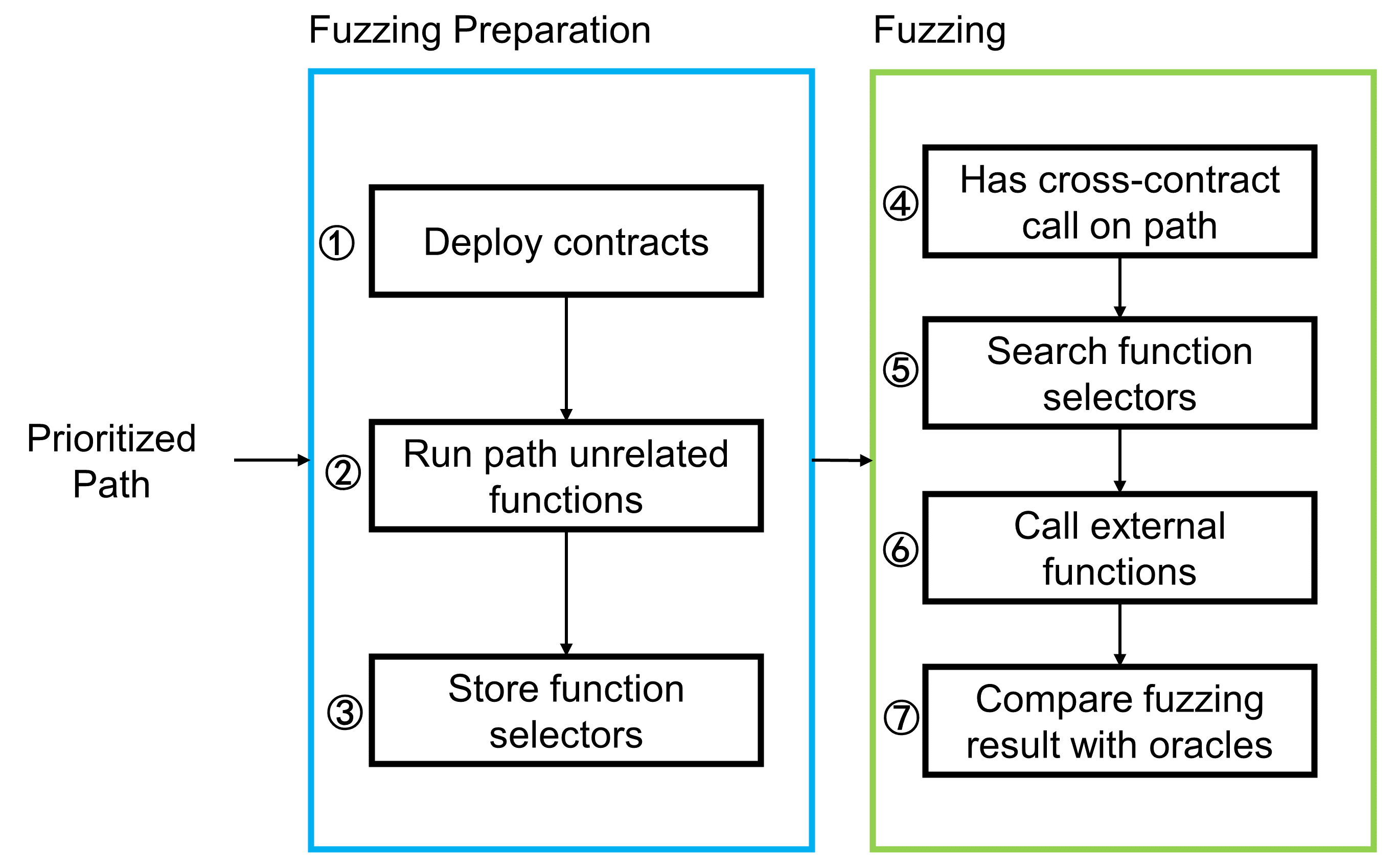}
	\caption{The cross-contract fuzzing process.}
	\label{fig:cross-contract fuzzing}
\end{figure}

Given the prioritized paths, we utilized cross-contract fuzzing to improve fuzzing efficiency. Here, we implement this fuzzing technique by the following steps: 1) The contracts under test should be deployed on EVM. As shown in Figure \ref{fig:cross-contract fuzzing}, the fuzzer will first deploy all contracts on a local private chain to facilitate cross-contract calls among contracts. 2) The path-unrelated functions will be called. Here, the path-unrelated functions denote functions that do not appear in the input prioritized paths. We run them first to initialize state variables of a contract. 3) We store the function selectors appeared in all contracts. The function selector is the unique identity recognizer of a function. It is usually encoded in 4-byte hex code~\cite{function-selector}. 4) The fuzzer checks whether there is a cross-contract call. If not, the following step 5 and step 6 will be skipped. 5) The fuzzer automatically searches local states to find out correct function selectors, and then directly trigger a cross-contract call to the target function in step 6. 7) The fuzzer compares the execution results against the detection rules and output reports. 

%% file: sec/6-eval.tex
\section{Evaluation}
\label{sec:evaluation}

\ourTool is implemented in Python and C with 3298 lines of code. All experiments are run on a computer which is running Ubuntu 18.04 LTS and equipped with Intel Xeon E5-2620v4, 32GB memories and 2TB HDD.

For the baseline comparison, \ourTool is compared with the state-of-art fuzzer \sfuzz~\cite{sfuzz}, a previously published testing engine \contractfuzzer~\cite{contractfuzzer} and a static cross-contract analysis tool~\clairvoyance \cite{clarivoyance}. The recently published tool \textsc{Echidna}~\cite{echidna} relies on manually written testing oracles, which may lead to different testing results depending on developer's expertise. Thus, it is not compared. Other tools (like \textsc{Harvey}~\cite{harvey}) are not publicly available for evaluation, and thus are not included in our evaluations. We systematically run all four tools on the contract datasets. Notably, to verify the authenticity of the vulnerability reports, we invite senior technical experts from security department of our industry partner to check vulnerable code. Our evaluation aims at investigating the following research questions (RQs).

\begin{enumerate}[leftmargin=*,label=\textbf{RQ\arabic*.},topsep=0pt,itemsep=0ex]
\item  How effective is \ourTool in detecting cross-contract vulnerabilities?
\item  To what extent the machine learning models and the path prioritization contribute to reducing the search space?
\item  What are the overhead of \ourTool, compared to the vanilla \sfuzz? 
\item  Can \ourTool discover real-world unknown cross-contract vulnerabilities, and what are the reasons for false negatives? 
\end{enumerate}


\subsection{Dataset Preparation}

Our evaluation dataset includes smart contracts from three sources: 1) datasets from previously published works (e.g., \cite{ren2021empirical} and \cite{smartbugs}); 2) smart contract vulnerability websites with good reputation (e.g., \cite{swcregis}); 3) smart contracts downloaded from Etherscan. The dataset is carefully checked to remove duplicate contracts with dataset used in our machine learning training. Specifically, the \textbf{DataSet1} includes contracts from previous works and famous websites. After we remove duplicate contracts and toy-contract (i.e., those which are not deployed on real world chains), we collect 18 labeled reentrancy vulnerabilities. To enrich the evaluation dataset, our \textbf{Dataset2} includes contracts downloaded from Etherscan. We remove contracts without external calls (they are irrelevant to cross-contract vulnerabilities) and contracts that are not developed by using Solidity 0.4.24 and 0.4.25 (i.e., the most two popular versions of Solidity~\cite{solidityversion}). In the end, \todo{7,391} contracts are collected in \textbf{Dataset2}. The source code of the above datasets are publicly available in our website~\cite{xFuzz} so that the evaluations are reproducible, benefiting further research.



\subsection{RQ1: Vulnerability Detection Effectiveness}

We first conduct evaluations on \textbf{Dataset1} by comparing three tools \textsc{ContractFuzzer}, \textsc{sFuzz} and \ourTool. The \textsc{Clairvoyance} is not included because it is a static analysis tool. For the sake of page space, we present a part of the results in Table \ref{tbl:open-dataset evaluation} with an overall summary and leave the whole list available at here\footnote{\url{https://anonymous.4open.science/r/xFuzzforReview-ICSE/Evaluation\%20on\%20Open-dataset.pdf}}.

In this evaluation, \textsc{ContractFuzzer} fail to find a vulnerability among the contracts. \textsc{sFuzz} missed \todo{3} vulnerabilities and outputted \todo{9} incorrect reports. Comparatively, \ourTool missed \todo{2} vulnerabilities and outputted \todo{6} incorrect reports. The reason of the missed vulnerabilities and incorrect reports lies on the difficult branch conditions (e.g., an \codeff{if} statement with 3 conditions) which blocks the fuzzer to traverse vulnerable branches. Note that \ourTool is equipped with model guidance so that it can focus on fuzzing suspicious functions and find more vulnerabilities than \textsc{sFuzz}.

\begin{table}[t]
\centering
\caption{Evaluations on Dataset1. The \ding{52} represents the tool successfully finds vulnerability in this function, otherwise the tool is marked with \ding{54}. }
\label{tbl:open-dataset evaluation}
\begin{tabular}{cccc}
\toprule
Address    & ContractFuzzer    & xFuzz                      & sFuzz                      \\
\midrule
0x7a8721a9 & \ding{54} & \ding{52} & \ding{54} \\
0x4e73b32e & \ding{54} & \ding{52} & \ding{52} \\
0xb5e1b1ee & \ding{54} & \ding{52} & \ding{52} \\
0xaae1f51c & \ding{54} & \ding{52} & \ding{52} \\
0x7541b76c & \ding{54} & \ding{52} & \ding{54} \\
...        & ...     & ...                        & ...       \\
\end{tabular}
\begin{tabular}{cccc}
\toprule
Summary  & ContractFuzzer & xFuzz & sFuzz \\
\midrule
     & 0/18 & 9/18 & 5/18 \\
\bottomrule
\end{tabular}
\end{table}

While we compare our tool with existing works on publicly available \textbf{Dataset1}, the dataset only provides non-cross-contract labels thus cannot be used to verify our detection ability on cross-contract ones. To complete this, we further evaluate the effectiveness of cross-contract and non-cross-contract fuzzing on \textbf{Dataset2}. To reduce the effect of randomness, we repeat each setting 20 times, and report the averaged results.

\input{table/rq1-cross}

\subsubsection{Cross-contract Vulnerability.} The results are summarized in Table \ref{tbl:rq1-cross}. Note that the ``P\%'' and ``R\%'' represent precision rate and recall rate, ``\#N'' is the number of vulnerability reports. ``C.V.'' means \clairvoyance and ``C.F.'' means \contractfuzzer. Cross-contract vulnerabilities are currently not supported by \contractfuzzer, \sfuzz and thus they report no vulnerabilities detected.


\textbf{Precision.} \clairvoyance managed to find \todo{7} true cross-contract reentrancy vulnerabilities. In comparison, \ourTool found \todo{9} cross-contract reentrancy, \todo{3} cross-contract delegatecall and \todo{2} cross-contract tx-origin vulnerabilities. The two tools found \todo{21} cross-contract vulnerabilities in total. \clairvoyance report \todo{16} vulnerabilities but only \todo{43.7\%} of them are true positives. In contrast, \ourTool generates \todo{18} (13+3+2) reports of three types of cross-contract vulnerabilities and all of them are true positives. The reason of the high false positive rate of \clairvoyance is mainly due to its static analysis based approach, without runtime validation. We further check the \todo{18} vulnerabilities on some third-party security expose websites~\cite{contract-library,swcregis,smartbugs} and we find \todo{15} of them are not flagged.

\textbf{Recall.} The \todo{9} vulnerabilities missed by \clairvoyance are all resulted from the abuse of detection rules, i.e., the vulnerable contracts are filtered out by unsound rules. In total, \todo{3} cross-contract vulnerabilities are missed by \ourTool. A close investigation shows that they are missed due to the complex path conditions, which blocks the input from penetrating the function. We also carefully check false negatives missed by \ourTool, and find they are not reported by \textsc{ConractFuzzer} and \textsc{sFuzz} as well. While existing works all fail to penetrate the complex path conditions, we believe this limitation can be addressed by future works.


\subsubsection{Non-Cross-contract Vulnerability.} 
The experiment results show that \ourTool improves detection of non-cross-contract vulnerabilities as well (see Table \ref{tbl:rq1-pr-rr}). 
For reentrancy, \contractfuzzer achieves the best \todo{100\%} precision rate but the worst \todo{1.7\%} recall rate. \sfuzz and \clairvoyance identified \todo{33.5\%} and \todo{40.4\%} vulnerabilities. \ourTool has a precision rate of 95.5\%, which is slightly lower than that of \contractfuzzer, and more importantly, the bests recall rate of 84.2\%. \ourTool exhibits strong capability in detecting vulnerabilities by finding a total of \todo{209 (149+35+25)} vulnerabilities.

\textbf{Precision.} For reentrancy, \clairvoyance reports \todo{75} false positives, because of \ding{182} the abuse of detection rules and \ding{183} unexpected jump to unreachable paths due to program errors. The \todo{11} false positives of \sfuzz are due to the misconceived ether transfer. \sfuzz captures ether transfers to locate dangerous calls. However, the ethers from attacker to victim is also falsely captured. The 7 false alarms of \ourTool are due to the mistakes of contract programmers by calling a nonexistent functions. These calls are however misconceived as vulnerabilities by \ourTool.



\input{table/rq1-tbl1}

\textbf{Recall.} \clairvoyance missed \todo{59.6\%} of the true positives. The root cause is the adoption of unsound rules during static analysis. \sfuzz missed 117 reentrancy vulnerabilities and \todo{16} delegatecall vulnerabilities due to \emph{(1)} timeout and \emph{(2)} incapability to find feasible paths to the vulnerability. \ourTool missed 27 vulnerabilities due to complex path conditions. 

\begin{tcolorbox}[size=title,rightrule=1mm, leftrule=1mm, toprule=0mm, bottomrule=0mm, arc=0pt,colback=gray!5,colframe=bleudefrance!75!black]
  { \textbf{Answer to RQ1: } Our tool \ourTool achieves a precision of 95.5\% and a recall of 84.6\%. Among the evaluated four methods, \ourTool achieves the best recall. Besides, \ourTool successfully finds \todo{209} real-world non-cross-contract vulnerabilities as well as \todo{18} real-world cross-contract vulnerabilities.}
\end{tcolorbox}

\begin{figure}[t]

	\centering
	\includegraphics[width=0.4\textwidth]{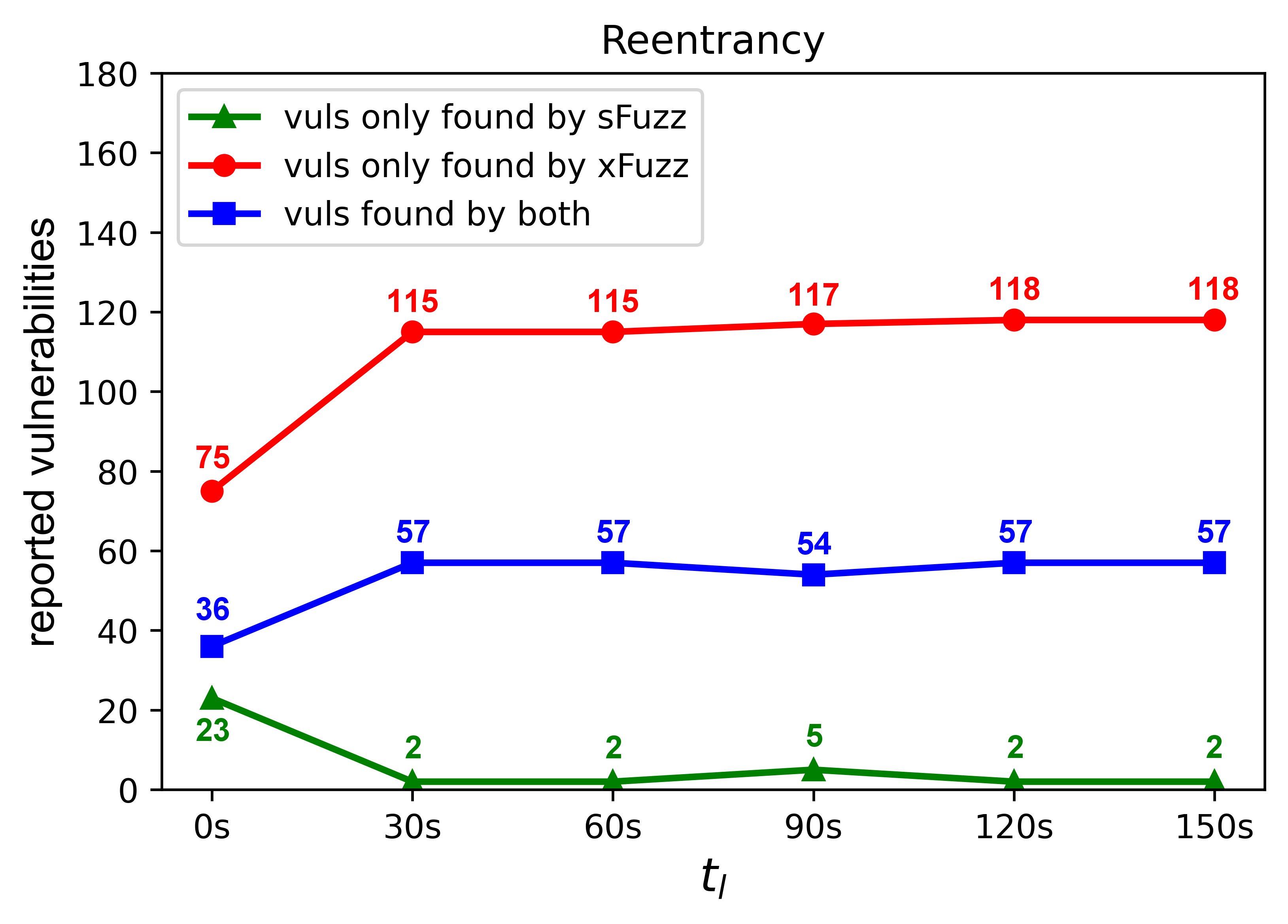}
	\caption{Comparison of reported vulnerabilities between \ourTool and \sfuzz regarding reentrancy.}  
	\label{fig:rq2-reentrancy}
\end{figure}

\begin{figure}[t]
	\centering
	\includegraphics[width=0.4\textwidth]{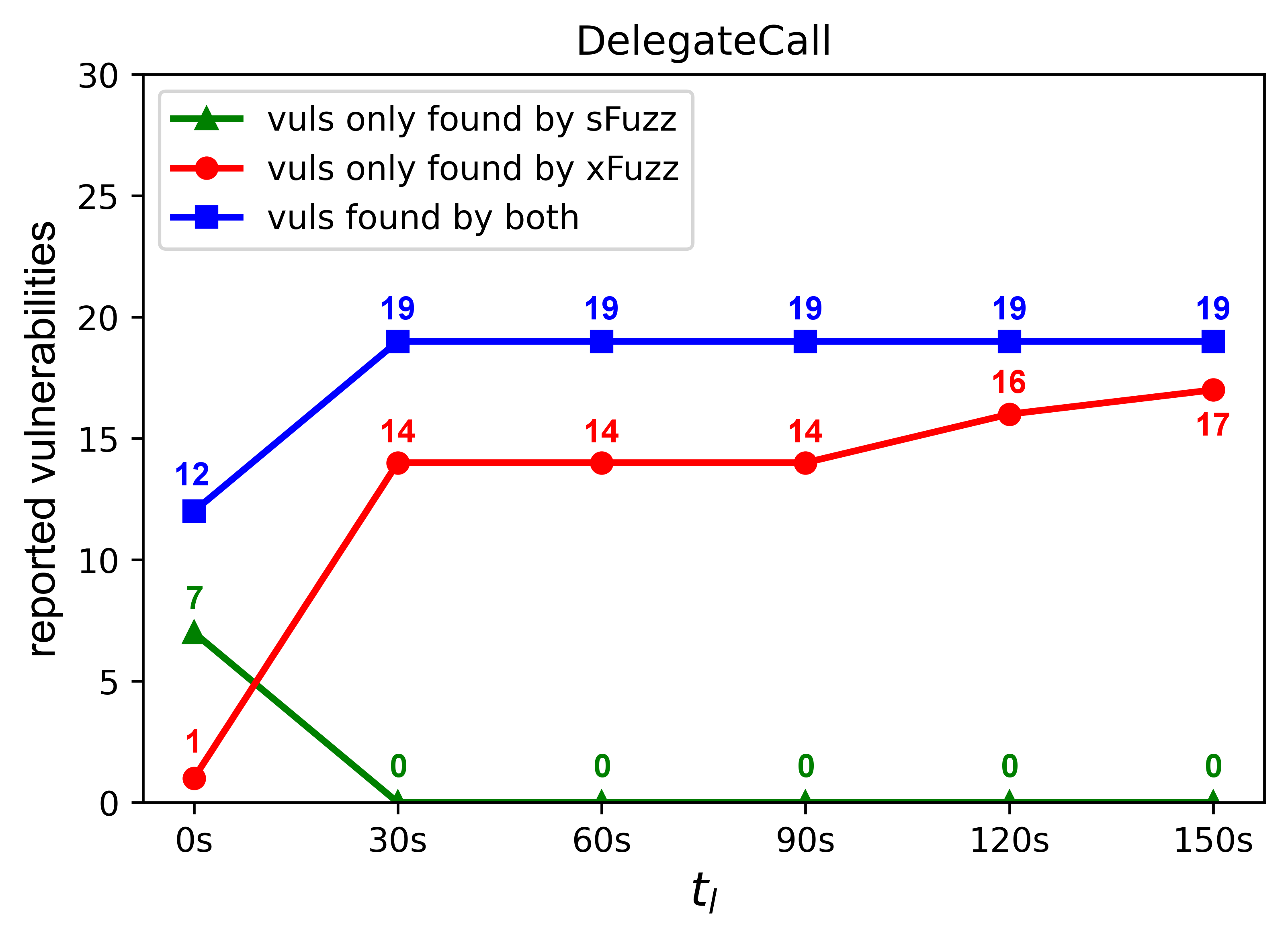}
	\caption{Comparison of reported vulnerabilities between \ourTool and \sfuzz regarding delegatecall.}
	\label{fig:rq2-delegate}
\end{figure}

\subsection{RQ2: The Effectiveness of Guided Testing}
This RQ investigates the usefulness of the ML model and path prioritization for the guidance of fuzzing. To answer this RQ, we compare \sfuzz with a customized version of \ourTool, i.e., which differs from \sfuzz only by adopting the ML model (without focusing on cross-contract vulnerabilities). The intuition is to check whether the ML model enables us to reduce the time spent on benign contracts and thus reveal vulnerabilities more efficiently. 
That is, we implement \ourTool such that each contract is only allowed to be fuzzed for $t_l$ seconds if the ML model considers the contract benign; or otherwise, 180 seconds, which is also the time limit adopted in \sfuzz. Note that if $t_l$ is 0, the contract is skipped entirely when it is predicted to be benign by the ML model. The goal is to see whether we can set $t_l$ to be a value smaller than 180 safely (i.e., without missing vulnerabilities). We thus systematically vary the value of $t_l$ and observe the number of identified vulnerabilities. 


The results are summarized in Figure \ref{fig:rq2-reentrancy} and Figure \ref{fig:rq2-delegate}. Note that the tx-origin vulnerability is not included since it is not supported by \sfuzz. The red line represents vulnerabilities only found by \ourTool, the green line represents vulnerabilities only reported by \sfuzz and the blue line denotes the reports shared by both two tools. We can see that the curves climb/drop sharply at the beginning and then saturate/flatten after 30s, indicating that most vulnerabilities are found in the first 30s. 

We observe that when $t_l$ is set to 0s (i.e., contracts predicted as benign are skipped entirely), \ourTool still detects \todo{82.8\%} (i.e., 111 out of 134, or equivalently 166\% of that of \sfuzz) of the reentrancy vulnerabilities as well as \todo{65.0\%} of the delegatecall vulnerability (13 out of 20). The result further improves if we set $t_l$ to be 30 seconds, i.e., almost all (except 2 out of 174 reentrancy vulnerabilities; and none of the delegatecall vulnerabilities) are identified. Based on the result, we conclude that the ML model indeed enables to reduce fuzzing time on likely benign contracts significantly (i.e., from 180 seconds to 30 seconds) without missing almost any vulnerability.



\textbf{The Effectiveness of Path Prioritization.} 
To evaluate the relevance of path prioritization, we further analyze the results of the customized version of \ourTool as discussed above. Recall that path prioritization allows us to explore likely vulnerable paths before the remaining. Thus, if path prioritization works, we would expect that the vulnerabilities are mostly found in paths, where \ourTool explores first. We thus systematically count the number of vulnerabilities found in the first 10 paths which are explored by \ourTool. The results are summarized in Table \ref{tbl:rq2-priority}, where column ``Top 10'' shows the number of vulnerabilities detected in the first 10 paths explored.
 
 

\input{table/rq2-priority}

The results show that, \ourTool finds a total of \todo{152} (out of 172) reentrancy vulnerabilities in the first 10 explored paths. In particular, the number of found vulnerabilities in the first 10 explored paths by \ourTool is almost three times as many as that by \sfuzz. Similarly, \ourTool also finds 32 (out of 33) delegatecall vulnerabilities in the first 10 explored paths. The results thus clearly suggest that path prioritization allows us to focus on relevant paths effectively, which has practical consequence on fuzzing large contracts. 


\begin{tcolorbox}[size=title,rightrule=1mm, leftrule=1mm, toprule=0mm, bottomrule=0mm, arc=0pt,colback=gray!5,colframe=bleudefrance!75!black]
  { \textbf{Answer to RQ2: } The ML model enables us to significantly reduce the fuzzing time on likely benign contracts without missing almost any vulnerabilities. Furthermore, most vulnerabilities are detected efficiently through our path prioritization. Overall, \ourTool finds \emph{twice} as many reentrancy or delegatecall vulnerabilities as \sfuzz.}
\end{tcolorbox}

\subsection{RQ3: Detection Efficiency}
\input{table/rq3-tbl3}
Next, we evaluate the efficiency of our approach. We record time taken for each step during fuzzing and the results are summarized in Table \ref{tbl:rq3-efficiency-3}. To eliminate randomness during fuzzing, we replay our experiments for five times and report the averaged results. In this table, ``MPT'' means model prediction time; ``ST'' means search time for vulnerable paths during fuzzing; ``DT'' means detection time for \clairvoyance and fuzzing time for the fuzzers. ``N.A.'' means that the tool has no such step in fuzzing or the vulnerability is currently not supported by it, and thus the time is not recorded.


The efficiency of our method (i.e., by reducing the search space) is evidenced as the results show that \ourTool is obviously faster than \sfuzz, i.e., saving \todo{80\%} of the time. The main reason for the saving is due to the saving on the search time (i.e., \todo{80\%} reduction).
We also observe that \ourTool is slightly slower than \sfuzz in terms of the effective fuzzing time, i.e., an additional \todo{32.5 (86.6-54.1)} min is used for fuzzing cross-contract vulnerabilities. This is expected as the number of paths is much more (even after the reduction thanks to the ML model and path prioritization) than that in the presence of more than 2 interacting contracts. Note that \clairvoyance is faster than all tools because this tool is a static detector without perform runtime execution of contracts.

\begin{tcolorbox}[size=title,rightrule=1mm, leftrule=1mm, toprule=0mm, bottomrule=0mm, arc=0pt,colback=gray!5,colframe=bleudefrance!75!black]
  { \textbf{Answer to RQ3: } Owing to the reduced search space of suspicious functions, the guided fuzzer \ourTool saves over 80\% of searching time and reports more vulnerabilities than \sfuzz with less than 20\% of the time.}
\end{tcolorbox}

\subsection{RQ4: Real-world Case Studies}
\label{sec: case study}


 \begin{figure}
    \centering
    \begin{minipage}[tb]{1.0\linewidth}
    \begin{lstlisting}
  function buyOne(address _exchange, uint256 _value, bytes _data) payable public
  {
    ...
    buyInternal(_exchange, _value, _data);
  }
  function buyInternal(address _exc, uint256 _value, bytes _data) internal
  {
    ...
    require(_exc.call.value(_value)(_data));
    balances[msg.sender] = balances[msg.sender].sub(_value);
  }
    \end{lstlisting}
    \caption{A real-world reentrancy vulnerability found by \ourTool, in which the vulnerable path relies on internal calls.}
    \label{fig:rq4-case2}	
    \end{minipage}
\end{figure}

In this section, we present \todo{2} typical vulnerabilities reported by \ourTool to qualitatively show why \ourTool works. In general, the ML model and path prioritization help \ourTool find vulnerabilities in three ways, i.e., \ding{182} locate vulnerable functions, \ding{183} identify paths from internal calls and \ding{184} identify feasible paths from external calls. 

 
\textbf{Real-world Case 1}: \ourTool is enhanced with path prioritization, which enables it to focus on vulnerabilities related to internal calls.
In Figure \ref{fig:rq4-case2}\footnote{deployed at 0x0695B9EA62C647E7621C84D12EFC9F2E0CDF5F72}, the modifier \codeff{internal} limits the access only to internal member functions. The attacker can however steal ethers by path \codeff{buyOne} $\rightarrow$ \codeff{buyInternal}. By applying \ourTool, the vulnerability is identified in 0.05 seconds and the vulnerable path is also efficiently exposed.

\textbf{Real-world Case 2}: The path prioritization also enables \ourTool to find cross-contract vulnerabilities efficiently. For example, a real-world cross-contract vulnerability\footnote{deployed at 0x165CFB9CCF8B185E03205AB4118EA6AFBDBA9203} is shown in Figure \ref{fig:rq4-case3}. This example is for auditing transactions in real-world and involves with over 2,000 dollars. In this example, function \codeff{registerAudit} has a cross-contract call to a public address \codeff{CSolidStamp} at line 13, which intends to forward the call to function \codeff{audContract}. While this function is only allowed to be accessed by the registered functions, as limited by modifier \codeff{onlyRegister}, we can bypass this restriction by a cross-contract call \codeff{registerAudit} $\rightarrow$ \codeff{audContrat}. Eventually, an attacker would be able to steals the ethers in seconds.

 \begin{figure}
    \centering
    \begin{minipage}[tb]{1.0\linewidth}
    \begin{lstlisting}
contract SolidStamp{
  function audContract(address _auditor) public onlyRegister
  {
    ...
    _auditor.transfer(reward.sub(commissionKept));
  }
}
contract SolidStampRegister{
  address public CSolidStamp;
  function registerAudit(bytes32 _codeHash) public
  {
    ...
    SolidStamp(CSolidStamp).audContract(msg.sender);
  }
}
    \end{lstlisting}
    \caption{A cross-contract vulnerability found by \ourTool. This contract is used in auditing transactions in real-world.}
    \label{fig:rq4-case3}	
    \end{minipage}
\end{figure}



\textbf{Real-world Case 4}:During our investigation on the experiment results, we gain the insights that \ourTool can be further improved in terms of handling complex path conditions. Complex path conditions often lead to prolonged fuzzing time or blocking penetration altogether. We identified a total of \todo{3} cross-contract and \todo{24} non-cross-contract vulnerabilities that are missed due to such a reason. Two of such complex condition examples (from two real-word false negatives of \ourTool) are shown in Figure \ref{fig:rq4-case4}. Function calls, values, variables and arrays are involved in the conditions. These conditions are difficult to satisfy for \ourTool and fuzzers in general (e.g., \sfuzz failed to penetrate these paths too). This problem can be potentially addressed by integrating \ourTool with a theorem prover such that Z3~\cite{z3} which is tasked to solve these path conditions. That is, a hybrid fuzzing approach that integrates symbolic execution in a lightweight manner is likely to further improve \ourTool.


 \begin{figure}
    \centering
    \begin{minipage}[htb]{1.0\linewidth}
    \begin{lstlisting}
if ((random()%2==1) && (msg.value == 1 ether) && (!locked))
\\at 0x11F4306f9812B80E75C1411C1cf296b04917b2f0

require(msg.value == 0 || (_amount == msg.value && etherTokens[fromToken]));
\\at 0x1a5f170802824e44181b6727e5447950880187ab
    \end{lstlisting}
    \caption{Complex path conditions involving with multiple variables and values.}
    \label{fig:rq4-case4}	
    \end{minipage}
\end{figure}


\begin{tcolorbox}[size=title,rightrule=1mm, leftrule=1mm, toprule=0mm, bottomrule=0mm, arc=0pt,colback=gray!5,colframe=bleudefrance!75!black]
  { \textbf{Answer to RQ4: } With the help of model predictions and path prioritization, \ourTool is capable of rapidly locating vulnerabilities in real-world contracts. The main reason for false negatives is complex path conditions, which could be potentially addressed through integrating hybrid fuzzing into \ourTool.}
\end{tcolorbox}

%% file: table/rq1-cross.tex
\begin{table}[t]
\footnotesize

\caption{Performance of \ourTool, \clairvoyance (C.V.), \contractfuzzer (C.F.), \sfuzz on cross-contract vulnerabilities.}
\label{tbl:rq1-cross}

\begin{tabular}{cccccccccc}
\toprule
\multirow{2}{*}{} & \multicolumn{3}{c}{reentrancy} & \multicolumn{3}{c}{delegatecall} & \multicolumn{3}{c}{tx-origin} \\ \cline{2-10} 
  & P\% & R\%  & \#N  & P\% & R\% & \#N & P\% & R\% & \#N  \\
\midrule
\textsc{C.F.} & 0 & 0 & 0 & 0 & 0 & 0 & 0 & 0 & 0  \\
\cellcolor{gray!15}\textsc{sFuzz} & \cellcolor{gray!15}0 & \cellcolor{gray!15}0 & \cellcolor{gray!15}0 & \cellcolor{gray!15}0 & \cellcolor{gray!15}0 & \cellcolor{gray!15}0 & \cellcolor{gray!15}0 & \cellcolor{gray!15}0 & \cellcolor{gray!15}0 \\
\textsc{C.V.}  & 43.7 & 43.7  & 16  & 0 & 0 & 0 & 0 & 0 & 0 \\ 
\cellcolor{gray!15}\textsc{xFuzz}  & \cellcolor{gray!15}100 & \cellcolor{gray!15}81.2 & \cellcolor{gray!15}13 & \cellcolor{gray!15}100 & \cellcolor{gray!15}100 & \cellcolor{gray!15}3  & \cellcolor{gray!15}100 & \cellcolor{gray!15}100  & \cellcolor{gray!15}2 \\
\bottomrule
\end{tabular}
\end{table}

%% file: table/rq1-tbl1.tex
\begin{table}[]
\footnotesize
\caption{Performance of \ourTool, \clairvoyance, \contractfuzzer and \sfuzz on non-cross-contract evaluations.}
\label{tbl:rq1-pr-rr}
\begin{tabular}{cccccccccc}
\toprule
\multirow{2}{*}{} & \multicolumn{3}{c}{reentrancy} & \multicolumn{3}{c}{delegatecall} & \multicolumn{3}{c}{tx-origin} \\ \cline{2-10} 
  & P\% & R\%  & \#N  & P\% & R\% & \#N & P\% & R\% & \#N  \\
\midrule
\textsc{C.F.} & 100 & 1.7 & 3 & 0 & 0 & 0 &  0 & 0 & 0  \\
\cellcolor{gray!15}\textsc{sFuzz} & \cellcolor{gray!15}\revf{84.2} & \cellcolor{gray!15}33.5 & \cellcolor{gray!15}{70} & \cellcolor{gray!15}100 & \cellcolor{gray!15}54.3 & \cellcolor{gray!15}19 & \cellcolor{gray!15}0 & \cellcolor{gray!15}0 & \cellcolor{gray!15}0 \\
\textsc{C.V.}  & 48.3  & 40.4   & 145  & 0 & 0 & 0 & 0 & 0 & 0  \\
\cellcolor{gray!15}\textsc{xFuzz}  & \cellcolor{gray!15}\revf{95.5} & \cellcolor{gray!15}\revf{84.6} & \cellcolor{gray!15}\revf{156} & \cellcolor{gray!15}100 & \cellcolor{gray!15}100 & \cellcolor{gray!15}35   & \cellcolor{gray!15}100 & \cellcolor{gray!15}100  & \cellcolor{gray!15}25  \\
\bottomrule
\end{tabular}
\end{table}

%% file: table/rq2-priority.tex
\begin{table}[t]
\centering
\caption{The paths reported by \ourTool and \sfuzz. The vulnerable paths found by the two tools are counted respectively.}
\label{tbl:rq2-priority}
\begin{tabular}{ccccc}
\toprule
\multirow{2}{*}{Found by} & \multirow{2}{*}{Vul} & \multirow{2}{*}{Total} & \multicolumn{2}{c}{Number in the Top} \\ \cline{4-5} 
      &              &     & Top10 & Other \\
\midrule
xFuzz & Reentrancy   & 172 & 152   & 20    \\ 
\cellcolor{gray!15}sFuzz & \cellcolor{gray!15}Reentrancy   & \cellcolor{gray!15}59  & \cellcolor{gray!15}57    & \cellcolor{gray!15}2     \\ 
xFuzz & Delegatecall & 33  & 32    & 1     \\ 
\cellcolor{gray!15}sFuzz & \cellcolor{gray!15}Delegatecall & \cellcolor{gray!15}19  & \cellcolor{gray!15}19    & \cellcolor{gray!15}0     \\ 
\bottomrule
\end{tabular}
\end{table}



%% file: table/rq3-tbl3.tex
\begin{table}[t]
\centering
\caption{The time cost of each step in fuzzing procedures.}
\label{tbl:rq3-efficiency-3}
\begin{tabular}{ccccc}
\toprule
\multicolumn{2}{c}{}          & sFuzz    & C.V. & xFuzz  \\ 
\midrule
\multirow{3}{*}{MPT(min)}   
  &  Reentrancy & N.A. &  N.A.    &  630.6  \\  
  & \cellcolor{gray!15}Delegatecall & \cellcolor{gray!15}N.A. & \cellcolor{gray!15}N.A.    & \cellcolor{gray!15}630.6  \\
  &  Tx-origin &  N.A.   &  N.A.    &  630.6  \\
\midrule
\multirow{3}{*}{ST(min)}    
  &  Reentrancy &  21,930.0   &  N.A.   &  3,621.0 \\  
  &  \cellcolor{gray!15}Delegatecall &  \cellcolor{gray!15}22,131.0   & \cellcolor{gray!15}N.A.  &  \cellcolor{gray!15}3,678.0  \\
  &  Tx-origin &  N.A. &  N.A.    &  3,683.0  \\
 \midrule
\multirow{3}{*}{DT(min)}    
    & Reentrancy & 54.1  & 246.2 & 86.6   \\ 
    & \cellcolor{gray!15}Delegatecall & \cellcolor{gray!15}2.8   & \cellcolor{gray!15}N.A.   & \cellcolor{gray!15}4.2   \\ 
    & Tx-origin & N.A.     & N.A.   &  2.9   \\ 
\midrule
\multirow{3}{*}{Total(min)} 
  & Reentrancy & 21,984.1   & 246.2   &  4,338.2  \\ 
  & \cellcolor{gray!15}Delegatecall & \cellcolor{gray!15}22,133.8   & \cellcolor{gray!15}N.A. & \cellcolor{gray!15}4,312.8  \\ 
  & Tx-origin & N.A.         & N.A.       &  4,316.5   \\
\bottomrule
\end{tabular}
\end{table}

%% file: sec/7-related.tex
\section{Related Work}
\label{sec:related work}

In this section, we discuss works that are most relevant to ours.

\textbf{Program analysis.} We draw valuable development experience and domain specific knowledge from existing work \cite{oyente, slither, vulnerability-survey-1, vulnerability-survey-2, vulnerability-survey-3}. Among them, \textsc{Slither} \cite{slither}, \textsc{Oyente} \cite{oyente} and Atzei \emph{et al.} \cite{vulnerability-survey-3} provide a transparent overlook on smart contracts detection and enhance our understanding on vulnerabilities. Chen \emph{et al.} \cite{vulnerability-survey-1} and Durieux \emph{et al.} \cite{vulnerability-survey-2} offer evaluations on the state-of-the-arts, which helps us find the limitation of existing tools.

\textbf{Cross-contract vulnerability.} Our study is closely related to previous works focusing on interactions between multiple contracts. Zhou \emph{et al.}~\cite{sasc} present work to analyze relevance between smart contract files, which inspires us to focus on cross-contract interactions. He \emph{et al.}~\cite{ilf} report that existing tools fail to exercise functions that can only execute at deeper states. Xue \emph{et al.}~\cite{clarivoyance} studied cross-contract reentrancy vulnerability. They propose to construct ICFG (combining CFGs with call graphs) then track vulnerability by taint analysis. 

\textbf{Smart contract testing.} Our study is also relevant to previous fuzzing work on smart contracts. Smart contract testing plays an important role in smart contract security. Zou \emph{et al.}~\cite{smart-contract-challenges} report that over 85\% of developers intend to do heavy testing when programming. The work of Jiang \emph{et al.} \cite{contractfuzzer} makes the early attempt to fuzz smart contracts. \textsc{ContractFuzzer} instruments Ethereum virtual machine and then collects execution logs for further analysis. Wüstholz \emph{et al.} present guided fuzzer to better mutate inputs. Similar method is implemented by He \emph{et al.}~\cite{ilf}. They propose to learn fuzzing strategies from the inputs generated from a symbolic expert. The above two methods inspire us to leverage a guider to reduce search space. Tai D \emph{et al.} \cite{sfuzz} implement a user-friendly AFL fuzzing tool for smart contracts, based on which we build our fuzzing framework. 
Different from these existing work, our work makes a special focus on proposing novel ML-guided method for fuzzing cross-contract vulnerabilities, which is highly important but largely untouched by existing work.
Additionally, our comprehensive evaluation demonstrates that our proposed technique indeed outperforms the state-of-the-arts in detecting cross-contract vulnerabilities.

\textbf{Machine learning practice.} This work is also inspired by previous work \cite{tzuyu, xue2015, exploitmeter}. In their work, they propose learning behavior automata to facilitate vulnerability detection. Zhuang \emph{et al.} \cite{smart-contract-graph-network} propose to build graph networks on smart contracts to extend understanding of malicious attacks. Their work inspires us to introduce machine learning method for detection. We also improve our model selection by inspiration of work of Liu \emph{et al.} \cite{easyensemble}. Their algorithm helps us select best models with satisfactory performance on recall and precision on highly imbalanced dataset. \revf{Yan \emph{et al.}~\cite{exploitmeter} have proposed a method to mimic the cognitive process of human experts. Their work inspires us to find the consensus of vulnerability evaluators to better train the machine learning models.} 

\textbf{Smart contract security to society.} Smart contract has drawn a number of security concerns since it came into being. As figured out by Zou \emph{et al.}~\cite{smart-contract-challenges}, over 75\% of developers agree that the smart contract software has a much high security requirement than traditional software. According to~\cite{smart-contract-challenges}, the reasons behind such requirement are: 1) The frequent operations on sensitive information (e.g., digital currencies, tokens); 2) The transactions are irreversible; 3) The deployed code cannot be modified. Considering the close connection between smart contract and financial activities, the security of smart contract security largely effects the stability of the society. 

%% file: sec/8-conclusion.tex
\section{Conclusion}
\label{sec:conclusion}

In this paper, we propose \ourTool, a novel machine learning guided fuzzing framework for smart contracts, with a special focus on cross-contract vulnerabilities. We address two key challenges during its development: the search space of fuzzing is reduced, and cross-contract fuzzing is completed. The experiments demonstrate that \ourTool is much faster and more effective than existing fuzzers and detectors. In future, we will extend our framework with more static approach to support more vulnerabilities.

%% file: sec/ref.tex
\bibliographystyle{IEEEtran}
\balance
\bibliography{ref}